\documentclass[twocolumn,showpacs,preprintnumbers,amsmath,amssymb,nofootinbib]{revtex4}

\usepackage{epsfig}
\usepackage{graphicx}
\usepackage{dcolumn}
\usepackage{bm}

\def\DESepsf(#1 width #2){\epsfxsize=#2 \epsfbox{#1}}

\begin{document}


\title{Charmless Two-body Baryonic $B$ Decays}

\author{Chun-Khiang Chua}

\affiliation{Department of Physics, National Taiwan University,
Taipei, Taiwan 10764, Republic of China }

\date{\today}

\begin{abstract}
We study charmless two-body baryonic $B$ decays in a diagramatic
approach.  Relations on decay amplitudes are obtained. In general
there are more than one tree and more than one penguin amplitudes.
The number of independent amplitudes can be reduced in the large
$m_B$ limit. It leads to more predictive results. Some prominent
modes for experimental searches are pointed out.
\end{abstract}


\maketitle

\section{Introduction}

Baryonic modes in $B$ decays are emerging. Decay modes, such as
$\bar B\to D^{*}\,p\,\bar n$~\cite{Anderson:2000tz},
$D^{*}\,p\,\bar p$~\cite{Abe:2002tw}, $p\,\bar p\,K^{(*)}$,
$p\overline p\,\pi$~\cite{Abe:2002ds,Wang:2003zy},
$\Lambda\overline p\pi^-$~\cite{Wang:2003yi} and
$\Lambda_c^+\overline p$~\cite{Gabyshev:2002dt}, have been
observed. The $\overline B {}^0\to\Lambda_c^+\overline p$ decay
having
\begin{equation}
{\mathcal B}(\overline B {}^0\to\Lambda_c^+\overline p)=
(2.19^{+0.56}_{-0.49}\pm0.32\pm0.57)\times 10^{-5},
\end{equation}
is the only two-body mode observed so
far~\cite{Abe:2002er,Gabyshev:2002dt}.
By a simple scaling of $|V_{ub}/V_{cb}|^2$~\cite{PDG} on the
$\overline B {}^0\to\Lambda_c^+\overline p$ decay rate, the rates
of the charmless modes are expected to be of the order $10^{-7}$
~\cite{Gabyshev:2002dt}. This estimation is consistent with the
(90\% confident level) experimental upper
limits~\cite{Abe:2002er},
\begin{eqnarray}
{\mathcal B}(\overline B {}^0\to p\overline p)<1.2\times 10^{-6},
\nonumber\\
{\mathcal B}(\overline B {}^0\to \Lambda\overline
\Lambda)<1.0\times 10^{-6},
\nonumber\\
{\mathcal B}(B^-\to \Lambda\overline p)<2.2\times 10^{-6},
\label{eq:UL}
\end{eqnarray}
obtained by using $31.7$ million $B\overline B$ events, and is
only one order of magnitude below. The number of $B\overline B$
samples are accumulating rapidly in B factories. The charmless
baryonic modes could be just around the corner.

Motivated by these observations, there are many recent theoretical
studies in the three-body decay
modes~\cite{Hou:2000bz,Chua:2001vh,Chua:2001xn,Cheng:2001tr,Chua:2002yd,Chua:2002wn,Cheng:2002fp,Rosner:2003bm,Arunagiri:2003hu}.
It is pointed out in Ref.~\cite{Hou:2000bz} that three-body
baryonic modes could be enhanced over two-body, by reducing energy
release to the baryons via emitting a fast recoil meson. The decay
rates of $D^{*}\,p\,\bar n$, $D^{*}\,p\,\bar p$, $p\,\bar p\,K$
and $\Lambda\overline p\pi^-$ modes can be understood to some
extent~\cite{Chua:2001vh,Chua:2001xn,Cheng:2001tr,Chua:2002yd,Chua:2002wn,Cheng:2002fp,Rosner:2003bm}
and the spectra, having threshold enhancement behavior, are
consistent with predictions~\cite{Hou:2000bz,Chua:2001xn}. The
three-body decays seem to be much involved than the two-body
decays. However, in some cases their amplitudes can be related to
some well measured quantities, such as nucleon magnetic form
factors, under the factorization approximation, and gain better
control.

The two-body baryonic decays are in general non-factorizable. One
has to resort to model calculations. There are pole
model~\cite{Deshpande:1987nc,Jarfi:1990ej,Cheng:2001tr,Cheng:2001ub},
sum rule~\cite{Chernyak:ag}, diquark
model~\cite{Ball:1990fw,Chang:2001jt} and flavor symmetry
related~\cite{Gronau:1987xq,Savage:jx,He:re,Sheikholeslami:fa,Sheikholeslami:vt,Luo:2003pv}
studies.
Predictions from various models usually differ a lot and early
calculations usually give too large rates. Some technics developed
may still be useful.
For example, an updated pole model prediction~\cite{Cheng:2001ub}
on the $\overline B {}^0\to\Lambda_c^+\overline p$ rate is
consistent with data~\cite{Gabyshev:2002dt}.

In this work we use a quark diagram (or topological) approach to
study charmless two-body baryonic $B$ decays. This approach was
developed and applied to the study of the two-body mesonic
decays~\cite{Zeppenfeld:1980ex,Chau:tk,Chau:1990ay,Gronau:1994rj,Gronau:1995hn,Cheng:2001sc}.
It is closely related to the SU(3) flavor
symmetry~\cite{Zeppenfeld:1980ex,Gronau:1994rj}. Furthermore, it
does not rely on any factorization assumption. It is stressed that
these topological amplitudes include long-distant and short
distant final state interaction (FSI)
effects~\cite{Chau:tk,Chau:1990ay}. For example, it is used in the
study on FSI in the $\overline B\to D P$
system~\cite{Cheng:2001sc}.

Motivated by the recent $\overline B {}^0\to \Lambda_c^+\overline
p$ observation, the topological approach has been applied to the
charmful baryonic case and the FSI effects are
studied~\cite{Luo:2003pv}.
We further extend the quark diagram approach to the charmless case
and obtain some amplitude relations. In general there are more
than one tree and more than one penguin amplitudes. In principle,
these amplitudes can be extracted from data. However, so far we do
not have any relevant data yet.

It is useful to reduce the number of independent topological
amplitudes. We use asymptotic relations~\cite{Brodsky:1980sx} to
relate various amplitudes. The same technics has been used in the
study of the three-body
case~\cite{Cheng:2001tr,Chua:2002wn,Chua:2002yd} and it leads to
encouraging results. For example, the experiment finding of
${\mathcal B}(\Lambda\overline p\pi^-)>{\mathcal
B}(\Sigma^0\overline p\pi^-)$~\cite{Wang:2003yi} can be
understood~\cite{Chua:2002yd} and tree-body decay spectra are
consistent with the QCD counting rule~\cite{Lepage:1979za}
expectations. Due to the large energy release, we expect the
asymptotic relations to work better in the two-body case than in
the three-body case. For example, the smallness of two-body decay
rates may due to some $1/m_B^2$ suppression as expected from QCD
counting rules.

The order of this paper is as follows. In
Sec.~\ref{sec:formulation}, we formulate the quark diagram
approach for the study of the charmless two-body baryonic $B$
decays. We consider decay modes with decuplet anti-decuplet, octet
anti-decuplet, decuplet anti-octet and octet anti-octet baryonic
final states. Relations on amplitudes are obtained. In
Sec.~\ref{sec:reduction}, we reduce the number of independent
topological amplitudes by considering the large $m_B$ limit. In
Sec.~\ref{sec:ph}, we discuss the phenomenology of the charmless
two-body baryonic decays. We suggest some prominent modes for
experimental searches. In Sec IV we give discussion and
conclusion, followed by an appendix for the derivation of
asymptotic relations.

\section{\label{sec:formulation}\boldmath Topological amplitudes of charmless $\overline B\to {\bf B}\overline{\bf B}$ decays}

In this section we use a quark diagram (or topological) approach
to decompose the charmless two-body baryonic decay amplitudes. It
is useful to re-derive some familiar results of the mesonic case
first. Since quark diagram is a representation of flavor SU(3)
symmetry, the topological approach should be closely related to
the SU(3) approach~\cite{Zeppenfeld:1980ex}. We use the $\overline
B \to D P$ decay to illustrate this point and to introduce some
useful tools before we turn to the charmless case.

We follow Ref.~\cite{Savage:ub} to decompose $\overline B \to D P$
decay amplitudes according to flavor SU(3) symmetry. We recall
that the fields annihilating $B^-,\,\overline B^0_{d,s}$, creating
$D^{0,+},\,D^+_s$ and creating $\pi,\,K,\,\eta_8$ transform
respectively as $\overline {\bold 3}$, ${\bold 3}$ and $\bold 8$
under SU(3)~\cite{Savage:ub,text},
\begin{eqnarray}
{\overline B}&=& \left(
\begin{array}{ccc}
B^- &\overline B {}^0 &\overline B {}^0_s
\end{array}
\right), \qquad
\overline D= \left(
\begin{array}{ccc}
\overline D^0 &D^- &D^-_s
\end{array}
\right), \nonumber\\
&&\Pi= \left(
\begin{array}{ccc}
{{\pi^0}\over\sqrt2}+{{\eta_8}\over\sqrt6}&{\pi^+} &{K^+}\\
{\pi^-}&-{{\pi^0}\over\sqrt2}+{{\eta_8}\over\sqrt6}&{K^0}\\
{K^-}  &{\overline K {}^0}&-\sqrt{2\over3}{\eta_8}
\end{array}
\right).
\end{eqnarray}
The $(\bar d u)(\bar c b)$ operators in the effective Hamiltonian
$H_{\rm W}$ can be expressed as $(\bar q_i H^i_j q^j)\,(\bar c
b)$, where $q^i=(u,\,d,\,s)$ and
\begin{equation}
H= \left(
\begin{array}{ccc}
0 &0 &0\\
1 &0 &0\\
0 &0 &0
\end{array}
\right).
\end{equation}
The effective Hamiltonian, in term of the meson degree of freedom,
for the $\overline B\to DP$ decay should have the same SU(3)
transform property of $H_{\rm W}$. Consequently, we
have~\cite{Savage:ub}
\begin{eqnarray}
H_{\rm eff}&=&T\, \overline B_m \overline D^m\, H^i_j\, \Pi^j_i
           +C\,\overline B_m \Pi^m_i\, H^i_j\,\overline D^j
\nonumber\\
           &&+E\,\overline B_i\,H^i_j\,\Pi^j_m \overline D^m,
\label{eq:TCE}
\end{eqnarray}
with probable FSI effects contained in the coefficients. The
$\overline B\to DP$ decay amplitudes can be expressed in terms of
these coefficients~\cite{Savage:ub}
\begin{eqnarray}
A_{D^0\pi^-}&=&{\it T}+{\it C}, \qquad A_{D^+\pi^-}={\it T}+{\it
E},
\nonumber \\
A_{D^0\pi^0}&=&\frac{1}{\sqrt2}(-{\it C}+{\it E}), \qquad A_{D^+_s
K^-}={\it E},
\nonumber \\
A_{D^0\eta_8}&=&\frac{1}{\sqrt6}({\it C}+{\it E}). \label{eq:Atop}
\end{eqnarray}
The above expression can also be obtained by using the topological
approach with the coefficients $T,\,C,\,E$ interpreted as the
(color-allowed) external $W$-emission, (color-suppressed) internal
$W$-emission and $W$-exchange tree amplitudes,
respectively~\cite{Chau:tk,Chau:1990ay,Gronau:1994rj,Gronau:1995hn,Cheng:2001sc}.

The one-to-one correspondence of the SU(3) parameters and the
topological amplitudes is not a coincidence. It can be understood
by using a flavor flow analysis. We take the first term of $H_{\rm
eff}$ for illustration. In $H_{\rm W}$ the decays are governed by
the $b\to c\, \bar q^j\, q_i $ transition with the corresponded
$H^i_j$ coupling. The first term of $H_{\rm eff}$ in
Eq.~(\ref{eq:TCE}) is $\overline B_m \overline D^m\,H^i_j\,
\Pi^j_i$. The $\overline B_m \overline D^m$ part can be
interpreted as a $\overline B_m$ to $D^m$ transition with the same
light anti-quark $\bar q_m$ ~$(\bar q^m)$~\footnote{We use
subscript and superscript according to the field convention. For
example, we assign a subscript (superscript) to the initial
(final) state anti-quark $\bar q_m$~($\bar q^m$).} flavor, while
the $\Pi^j_i$ part is responsible for the creation of the meson
where the $W$-emitted $\bar q^j q_i$ pair ends up with. The above
picture clearly corresponds to the external $W$-emission topology.
Similarly, the identification of the $C$ ($E$) amplitude to the
second (third) term of $H_{\rm eff}$ can be understood in the same
way.

It is straightforward to extend the above approach to the
charmless case and the well known topological decompositions of
the charmless $\overline B\to PP$ decay
amplitudes~\cite{Chau:1990ay,Gronau:1994rj,Gronau:1995hn} can be
reproduced. For the $b\to u\bar u d$ and $b\to q\bar q d$
processes, the tree (${\cal O}_T$) and penguin (${\cal O}_P$)
operators respectively have the following flavor quantum numbers
\begin{eqnarray}
&&{\cal O}_T\sim (b \bar u)(u \bar d)
 =H^{ik}_j (b\bar q_i) (q^j\bar q_k),
\nonumber\\
&&{\cal O}_P\sim(b \bar q_i) (q^i \bar d)
 =H^k (b \bar q_i) (q^i \bar q_k),
\end{eqnarray}
with $H^{12}_1=1=H^2$, otherwise $H^{ik}_j=H^k=0$. The flavor
structures of $|\Delta S|=1$ tree and penguin operators can be
obtained by replacing $d$ to $s$ and $H^{12}_1=1=H^2$ to
$H^{13}_1=1=H^3$ in the above expression. By using a similar
flavor flow analysis as the $\overline B\to D P$ case, we
obtain~\footnote{Note that $H^{ik}_i(=H^k)$ does not lead to any
additional term.}
\begin{eqnarray}
H_{\rm eff}&=&T\, \overline B_m H^{ik}_j \Pi^j_k \Pi^m_i
              +C\,\overline B_m H^{ik}_j \Pi^j_i \Pi^m_k
\nonumber\\
              &&+E\,\overline B_k H^{ik}_j \Pi^j_l \Pi^l_i
              +A\,\overline B_i H^{ik}_j \Pi^j_l \Pi^l_k
\nonumber\\
            &&+P\,\overline B_m H^k \Pi^m_i \Pi^i_k
              +\frac{1}{2}PA\,\overline B_k H^k \Pi^l_m
              \Pi^m_l,
 \label{eq:B2PP}
\end{eqnarray}
where the $A$, $P$ and $PA$ terms correspond to annihilation,
penguin and penguin annihilation amplitudes, respectively. We can
reproduce Table I and II of Ref.~\cite{Gronau:1994rj} (up to some
trivial overall sign changes from wave function definitions) by
using the above $H_{\rm eff}$.

We are now ready to turn to the baryonic cases. We will study
various decay modes, including decuplet anti-decuplet, decuplet
anti-octet, octet anti-decuplet and octet anti-octet baryonic
final states. We start from the easiest case, in the sense of
flavor structure, and move on with increasing complexity.

\subsection{\boldmath $\overline B$ to decuplet anti-decuplet baryonic decays}

\begin{figure}[t!]
\centerline{\DESepsf(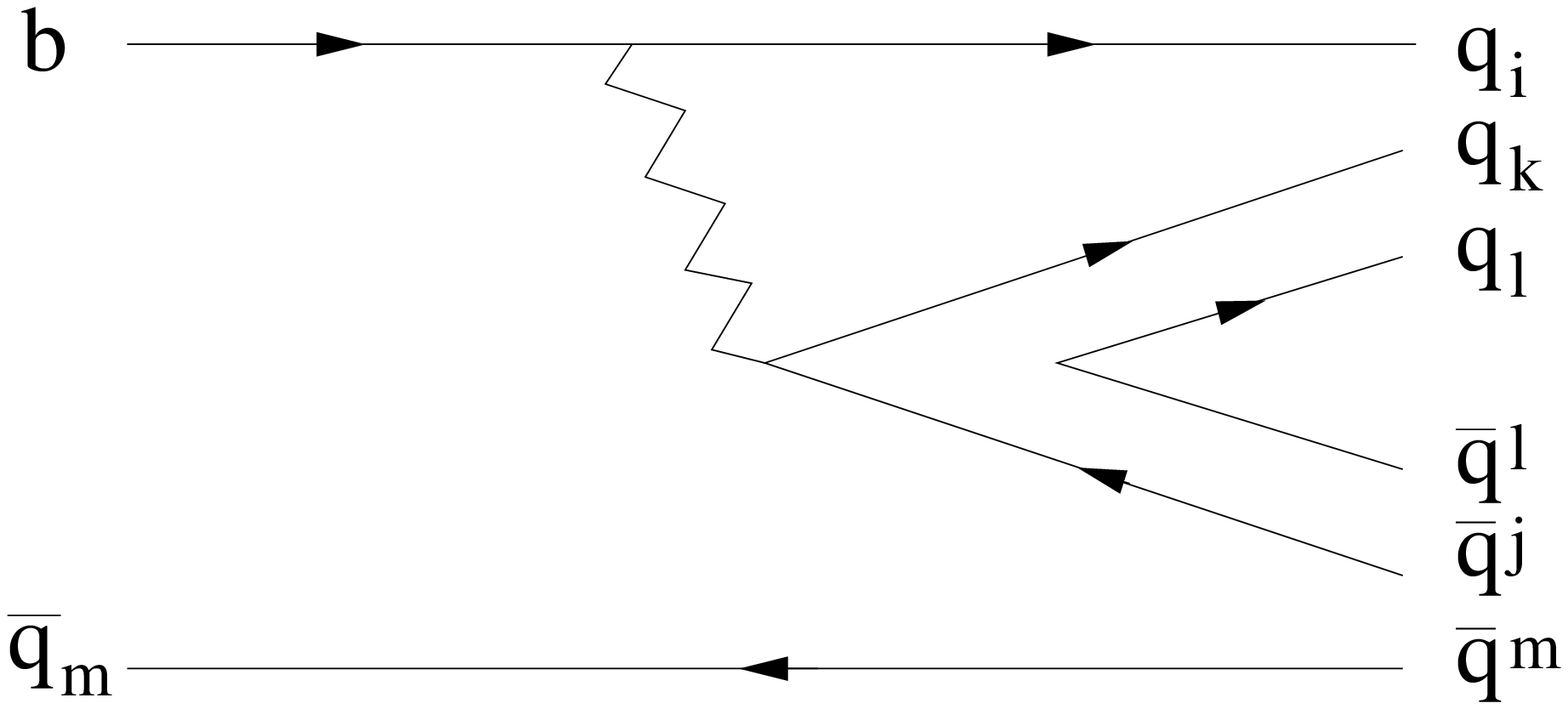 width 7.2cm)}
\centerline{(a)} \vskip0.2cm
\smallskip
\centerline{\DESepsf(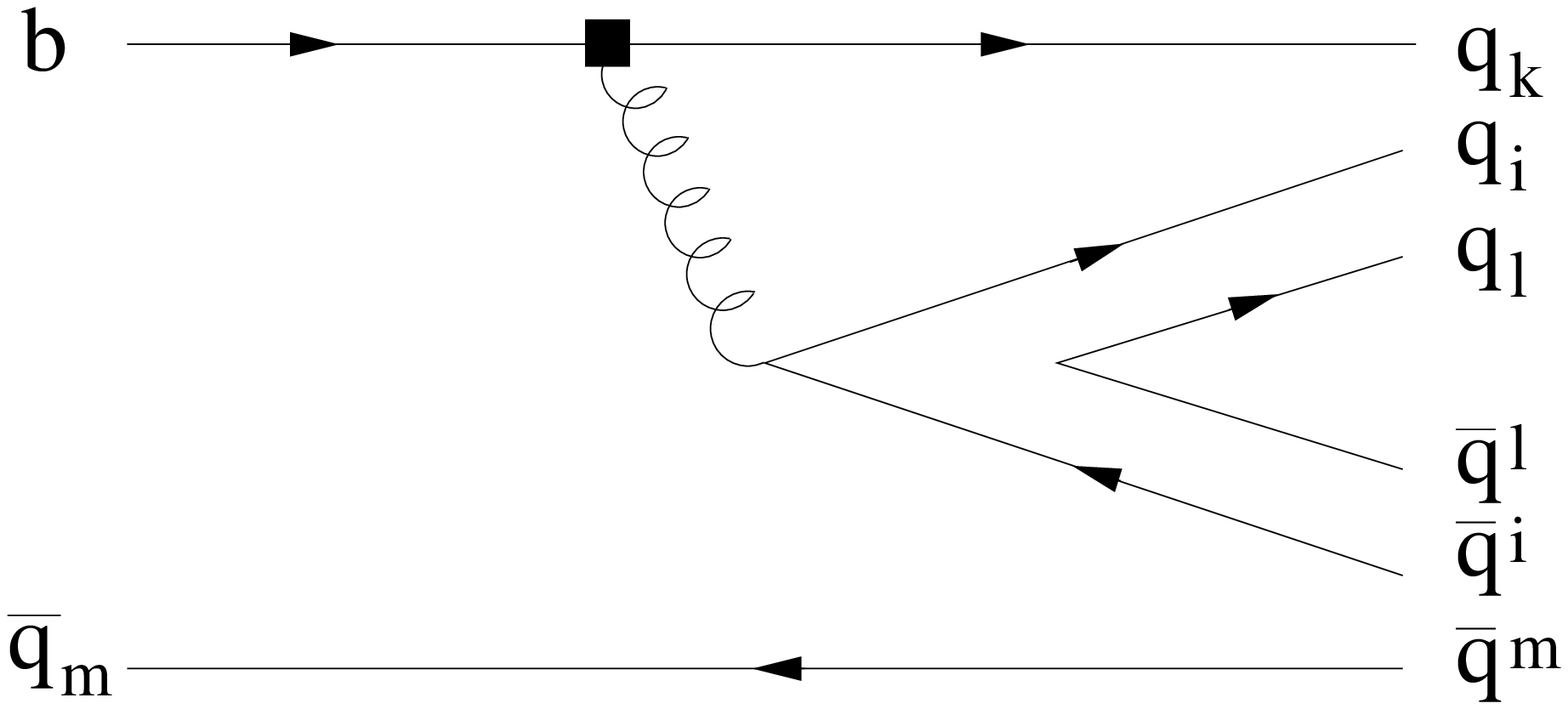 width 7.2cm)}
\centerline{(b)} \vskip0.2cm
\smallskip
\centerline{\DESepsf(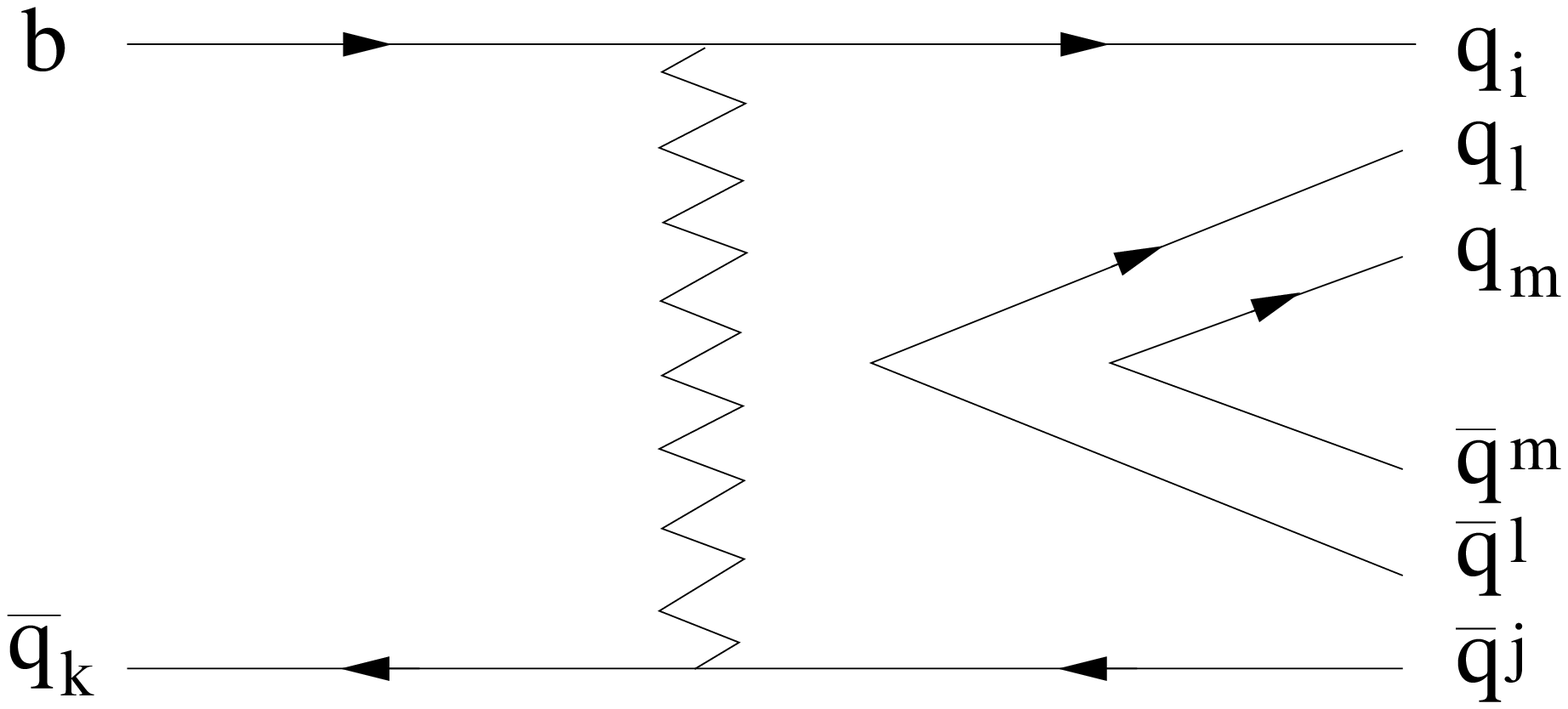 width 7.2cm)} \centerline{(c)}
\smallskip
\caption{Pictorial representation of
  (a) $T$ (tree), (b) $P$ (penguin) and (c) $E$ ($W$-exchange)
  amplitudes in $\overline B$ to baryon pair decays.
} \label{fig:TPE}
\end{figure}

It is straightforward to extent the quark diagram approach to the
$B$ to decuplet anti-decuplet decay. As shown in
Fig.~\ref{fig:TPE}, we have tree ($T$), penguin ($P$) and
$W$-exchange ($E$) amplitudes for processes governed by ${\cal
O}_{T,P}$.
Since there is no external $W$-exchange diagram and the internal
$W$-exchange amplitude is not color suppressed, we use the symbol
$T$ for the tree amplitude.

The decomposition of decay amplitudes can be obtained by using the
flavor flow analysis. The first step is to match the flavor. For
example, the decuplet with $q_i q_k q_l$ flavor as shown in
Fig.~\ref{fig:TPE}(a) is produced by the $\overline {\cal
D}_{ikl}$ field, while the decuplet with $\bar q^l \bar q^j \bar
q^m$ flavor is created by the ${\cal D}^{jlm}$ field, where ${\cal
D}^{jlm}$ is the familiar decuplet field with symmetric flavor
indices. To be specified, we use (see, for example~\cite{text})
\begin{eqnarray}
 &&{\cal D}^{111}=\Delta^{++},\,\,
 {\cal D}^{112}=\frac{1}{\sqrt3}\Delta^{+},
 \nonumber\\
 &&{\cal D}^{122}=\frac{1}{\sqrt3}\Delta^{0},\,\,
 {\cal D}^{222}=\Delta^{-},
 \nonumber\\
 &&{\cal D}^{113}=\frac{1}{\sqrt3}\Sigma^{*+},\,\,
 {\cal D}^{123}=\frac{1}{\sqrt6}\Sigma^{*0},\,\,
 {\cal D}^{223}=\frac{1}{\sqrt3}\Sigma^{*-},\,\,
 \nonumber\\
 &&{\cal D}^{133}=\frac{1}{\sqrt3}\Xi^{*0},\,\,
{\cal D}^{233}=\frac{1}{\sqrt3}\Xi^{*-},\,\,
 {\cal D}^{333}=\Omega^{-}.
\end{eqnarray}
Hence by using the flavor flow analysis and the corresponding
rule,
\begin{equation} q_i q_k
q_l\to\overline {\cal D}_{ikl},\qquad
  \bar q^l \bar q^j \bar q^m\to{\cal D}^{ljm},
\end{equation}
we have
\begin{eqnarray}
H_{\rm eff}&=& 6\,T_{{\cal D}\overline{\cal D}}\,\overline B_m
H^{ik}_j \overline {\cal D}_{ikl} {\cal D}^{ljm}
              +6\,P_{{\cal D}\overline{\cal D}}\,\overline B_m H^k
             \overline {\cal D}_{kil} {\cal D}^{lim}
\nonumber\\  &&+E_{{\cal D}\overline{\cal D}}\,\overline B_k
                H^{ik}_j
              \overline {\cal D}_{ilm} {\cal D}^{mlj}
             +A_{{\cal D}\overline{\cal D}}\,\overline B_i H^{ik}_j
               \overline {\cal D}_{klm} {\cal D}^{mlj}
\nonumber\\
            && +PA_{{\cal D}\overline{\cal D}}\,\overline B_k H^k
               \overline {\cal D}_{lmn} {\cal D}^{nml}.
              \label{eq:DD}
\end{eqnarray}
The pre-factors before $T_{{\cal D}\overline{\cal D}}$ and
$P_{{\cal D}\overline{\cal D}}$ are assigned for latter purpose.
The above equation is an extension of Eq.~(\ref{eq:B2PP}).

It is useful to discuss the QCD counting rules for these
amplitudes. In the large $m_B$ limit, we need a hard gluon to
create a $q\bar q$ pair for the tree or penguin topology. An
additional gluon is required to kick the spectator quark in the
$B$ meson such that it becomes energetic in the final baryon pair.
In the large $m_B$ limit, these amplitudes behave like $\sim
1/m_B^4$. The $1/m_B^4$ factor and the Cabibbo-Kobayashi-Maskawa
(CKM) coefficients lead to a suppressed charmless two-body
baryonic rates that may be the underlying reason of the negative
search result. Other topologies, such as $W$-exchange ($E_{{\cal
D}\overline{\cal D}}$), weak-annihilation $A_{{\cal
D}\overline{\cal D}}$ and penguin-annihilation ($PA_{{\cal
D}\overline{\cal D}}$) topologies, require an additional $q\bar q$
pair creation (see, for example Fig.~\ref{fig:TPE}(c)) and are
further $1/m_B^2$ suppressed. We should concentrate on $T$ and $P$
amplitudes. The decomposition of $\overline B$ to decuplet
anti-decuplet decay amplitudes are given in Table~\ref{tab:DD}. We
use $T^{(\prime)}$ and $P^{(\prime)}$ for the tree and penguin
amplitudes, respectively, in $|\Delta S|=0(1)$ processes.

\begin{table}[b!]
\caption{\label{tab:DD} Decomposition of $\overline B\to {\cal
D}\overline{\cal D}$ amplitudes in terms of tree and penguin
amplitudes.}
\begin{ruledtabular}
\begin{tabular}{rccrcc}
Mode
          & $T_{{\cal D}\overline{\cal D}}$
          & $P_{{\cal D}\overline{\cal D}}$
          & Mode
          & $T^\prime_{{\cal D}\overline{\cal D}}$
          & $P^\prime_{{\cal D}\overline{\cal D}}$
          \\
\hline $B^-\to \Delta^{+} \overline{\Delta^{++}}$
          & $2\sqrt{3}$
          & ${2}{\sqrt3}$
          & $B^-\to \Sigma^{*+} \overline{\Delta^{++}}$
          & $2\sqrt{3}$
          & ${2}{\sqrt3}$
          \\
$\Delta^0 \overline{\Delta^+}$
          & $2$
          & ${4}$
          & $\Sigma^{*0} \overline{\Delta^+}$
          & $\sqrt2$
          & ${2\sqrt2}$
          \\
$\Sigma^{*0} \overline{\Sigma^{*+}}$
          & $\sqrt2$
          & ${2\sqrt2}$
          &$\Xi^{*0} \overline{\Sigma^{*+}}$
          & $ 2$
          & ${4}$
          \\
$\Sigma^{*-} \overline{\Sigma^{*0}}$
          & 0
          & ${2\sqrt2}$
          & $\Xi^{*-} \overline{\Sigma^{*0}}$
          & 0
          & ${2\sqrt2}$
          \\
$\Xi^{*-} \overline{\Xi^{*0}}$
          & 0
          & ${2}$
          & $\Omega^{-} \overline{\Xi^{*0}}$
          & 0
          & ${2}{\sqrt3}$
          \\
$\Delta^- \overline{\Delta^0}$
          & 0
          & ${2}{\sqrt3}$
          & $\Sigma^{*-} \overline{\Delta^0}$
          & 0
          & ${2}$
          \\
\hline $\overline B {}^0\to \Delta^+ \overline{\Delta^{+}}$
          & $2$
          & ${2}$
          & $\overline B {}^0\to \Sigma^{*+} \overline{\Delta^{+}}$
          & $2$
          & ${2}$
          \\
$\Delta^0 \overline{\Delta^0}$
          & 2
          & ${4}$
          & $\Sigma^{*0} \overline{\Delta^0}$
          & $\sqrt2$
          & ${2\sqrt2}$
          \\
$\Sigma^{*0} \overline{\Sigma^{*0}}$
          & 1
          & ${2}$
          & $\Xi^{*0} \overline{\Sigma^{*0}}$
          & $\sqrt2$
          & ${2\sqrt2}$
          \\
$\Sigma^{*-} \overline{\Sigma^{*-}}$
          & 0
          & ${4}$
          & $\Xi^{*-} \overline{\Sigma^{*-}}$
          & 0
          & ${4}$
          \\
$\Xi^{*-} \overline{\Xi^{*-}}$
          & 0
          & ${2}$
          & $\Omega^{-} \overline{\Xi^{*-}}$
          & 0
          & ${2}{\sqrt3}$
          \\
$\Delta^- \overline{\Delta^-}$
          & $0$
          & $6$
          & $\Sigma^{*-} \overline{\Delta^-}$
          & $0$
          & ${2}{\sqrt3}$
          \\
\hline $\overline B {}^0_s\to \Delta^{+} \overline{\Sigma^{*+}}$
          & 2
          & ${2}$
          & $\overline B {}^0_s\to \Sigma^{*+} \overline{\Sigma^{*+}}$
          & 2
          & ${2}$
          \\
$\Delta^0 \overline{\Sigma^{*0}}$
          & $\sqrt2$
          & ${2\sqrt2}$
          & $\Sigma^{*0} \overline{\Sigma^{*0}}$
          & $1$
          & ${2}$
          \\
$\Sigma^{*0} \overline{\Xi^{*0}}$
          & $\sqrt2$
          & ${2\sqrt2}$
          & $\Xi^{*0} \overline{\Xi^{*0}}$
          & $2$
          & ${4}$
          \\
$\Sigma^{*-} \overline{\Xi^{*-}}$
          & $0$
          & ${4}$
          & $\Xi^{*-} \overline{\Xi^{*-}}$
          & $0$
          & ${4}$
          \\
$\Xi^{*-} \overline{\Omega^{-}}$
          & 0
          & ${2}{\sqrt3}$
          & $\Omega^{-} \overline{\Omega^{-}}$
          & 0
          & $6$
          \\
$\Delta^{-} \overline{\Sigma^{*-}}$
          & 0
          & ${2}{\sqrt3}$
          & $\Sigma^{*-} \overline{\Sigma^{*-}}$
          & 0
          & ${2}$
          \\
\end{tabular}
\end{ruledtabular}
\end{table}

By using the decay amplitudes shown in Table~\ref{tab:DD} we
obtain some amplitude relations. For the $\Delta S=0$ case, we
have
\begin{widetext}
\begin{eqnarray}
 A(B^-\to\Delta^+\overline{\Delta^{++}})
 &-&\sqrt3\,A(B^-\to\Delta^0\overline{\Delta^{+}})
+A(B^-\to\Delta^-\overline{\Delta^{0}})
 =0,
 \nonumber\\
A(B^-\to\Delta^+\overline{\Delta^{++}})
 &=&\sqrt3\, A(\overline B {}^0\to \Delta^+\overline{\Delta^+})
 =\sqrt3\, A(\overline B {}^0_s\to \Delta^+\overline{\Sigma^{*+}}),
 \nonumber\\
A(B^-\to\Delta^0\overline{\Delta^{+}})
 &=&A(\overline B {}^0\to \Delta^0\overline{\Delta^0})
 =\sqrt2\, A(\overline B {}^0_s\to \Delta^0\overline{\Sigma^{*0}})
 \nonumber\\
&=&\sqrt2\,A(B^-\to\Sigma^{*0}\overline{\Sigma^{*+}})
 =2\, A(\overline B {}^0\to \Sigma^{*0}\overline{\Sigma^{*0}})
 =\sqrt2\,A(\overline B {}^0_s\to \Sigma^{*0}\overline{\Xi^{*0}}),
 \nonumber\\
2\sqrt3\,A(B^-\to\Delta^{-}\overline{\Delta^{0}})
 &=&2\, A(\overline B {}^0\to \Delta^{-}\overline{\Delta^{-}})
 =2\sqrt3\,A(\overline B {}^0_s\to \Delta^{-}\overline{\Sigma^{*-}})
\nonumber\\
&=&3\sqrt2\,A(B^-\to\Sigma^{*-}\overline{\Sigma^{*0}})
 =3\, A(\overline B {}^0\to \Sigma^{*-}\overline{\Sigma^{*-}})
 =3\,A(\overline B {}^0_s\to \Sigma^{*-}\overline{\Xi^{*-}})
 \nonumber\\
&=&6\,A(B^-\to\Xi^{*-}\overline{\Xi^{*0}})
 =6\, A(\overline B {}^0\to \Xi^{*-}\overline{\Xi^{*-}})
 =2\sqrt3\,A(\overline B {}^0_s\to \Xi^{*-}\overline{\Omega^{-}}),
 \label{eq:relDD0}
\end{eqnarray}
and for the $|\Delta S|=1$ case, we have
\begin{eqnarray}
 A(B^-\to\Sigma^{*+}\overline{\Delta^{++}})
   &-&\sqrt6\,A(B^-\to\Sigma^{*0}\overline{\Delta^{+}})
   +\sqrt3\,A(B^-\to\Sigma^{*-}\overline{\Delta^{0}})
   =0,
\nonumber\\
A(B^-\to\Sigma^{*+}\overline{\Delta^{++}})
 &=&\sqrt3\, A(\overline B {}^0\to \Sigma^{*+}\overline{\Delta^+})
 =\sqrt3\, A(\overline B {}^0_s\to \Sigma^{*+}\overline{\Sigma^{*+}}),
 \nonumber\\
\sqrt2\,A(B^-\to\Sigma^{*0}\overline{\Delta^{+}})
 &=&\sqrt2\,A(\overline B {}^0\to \Sigma^{*0}\overline{\Delta^0})
 =2\, A(\overline B {}^0_s\to \Sigma^{*0}\overline{\Sigma^{*0}})
 \nonumber\\
\quad &=&A(B^-\to\Xi^{*0}\overline{\Sigma^{*+}})
 =\sqrt2\, A(\overline B {}^0\to \Xi^{*0}\overline{\Sigma^{*0}})
 =A(\overline B {}^0_s\to \Xi^{*0}\overline{\Xi^{*0}}),
 \nonumber\\
6\,A(B^-\to\Sigma^{*-}\overline{\Delta^{0}})
 &=&2\sqrt3\, A(\overline B {}^0\to \Sigma^{*-}\overline{\Delta^{-}})
 =6\,A(\overline B {}^0_s\to \Sigma^{*-}\overline{\Sigma^{*-}})
\nonumber\\
 &=&3\sqrt2\,A(B^-\to\Xi^{*-}\overline{\Sigma^{*0}})
 =3\, A(\overline B {}^0\to \Xi^{*-}\overline{\Sigma^{*-}})
 =3\,A(\overline B {}^0_s\to \Xi^{*-}\overline{\Xi^{*-}})
 \nonumber\\
&=&2\sqrt3\,A(B^-\to\Omega^{-}\overline{\Xi^{*0}})
 =2\sqrt3\, A(\overline B {}^0\to \Omega^{-}\overline{\Xi^{*-}})
 =2\,A(\overline B {}^0_s\to \Omega^{-}\overline{\Omega^{-}}).
  \label{eq:relDD1}
\end{eqnarray}
\end{widetext}
These relations are consistent with Ref.~\cite{Sheikholeslami:vt}.
Note that some results in Ref.~\cite{Sheikholeslami:vt} are
obtained by considering dominant tree or penguin amplitudes only,
while the above relations include both contributions.

\subsection{\boldmath $\overline B$ to decuplet anti-octet, octet anti-decuplet baryonic decays}

We extend the previous case to the $\overline B\to {\cal
B}\overline {\cal D}$, ${\cal D}\overline {\cal B}$ cases. Note
that the (anti-)decuplet parts are as before. For the octet part,
we use~\cite{text}
\begin{eqnarray}
{\mathcal B}= \left(
\begin{array}{ccc}
{{\Sigma^0}\over\sqrt2}+{{\Lambda}\over\sqrt6}
       &{\Sigma^+}
       &{p}
       \\
{\Sigma^-}
       &-{{\Sigma^0}\over\sqrt2}+{{\Lambda}\over\sqrt6}
       &{n}
       \\
{\Xi^-}
       &{\Xi^0}
       &-\sqrt{2\over3}{\Lambda}
\end{array}
\right). \label{eq:octet}
\end{eqnarray}
The ${\cal B}^j_k$ has a flavor structure $q^j q^a q^b
\epsilon_{abk}-\frac{1}{3}\,\delta^j_k q^c q^a q^b$~\cite{text}.
To match the $q_i q_k q_l$, $\bar q^l\bar q^j\bar q^m$ flavor
contents of final state octet baryons (as shown in
Fig.~\ref{fig:TPE}), we use
\begin{eqnarray}
q_i q_k q_l&\to& \epsilon_{ika} \overline {\cal B}^a_l,\,\
                 \epsilon_{ial} \overline {\cal B}^a_k,\,\
                 \epsilon_{akl} \overline {\cal B}^a_i,
\nonumber\\
\bar q^l\bar q^j\bar q^m&\to& \epsilon^{ljb} {\cal B}^m_b,\,\
                              \epsilon^{lbm} {\cal B}^j_b,\,\
                              \epsilon^{bjm} {\cal B}^l_b,
\label{eq:qqq}
\end{eqnarray}
as the corresponding fields in $H_{\rm eff}$.
In fact, not all terms in the above equation are independent. They
are constrained by
\begin{equation}
  \epsilon_{ika} \overline {\cal B}^a_l
 +\epsilon_{ial} \overline {\cal B}^a_k
 +\epsilon_{akl} \overline {\cal B}^a_i
 =0=
  \epsilon^{ljb} {\cal B}^m_b
 +\epsilon^{lbm} {\cal B}^j_b
 +\epsilon^{bjm} {\cal B}^l_b,
\label{eq:identity}
\end{equation}
which can be shown easily. Hence for each of the $q_i q_k q_l$ and
$\bar q^l\bar q^j\bar q^m$ configuration we only need two
independent terms.

\begin{table*}[t!]
\caption{\label{tab:BD} Decomposition of $\overline B\to {\cal
B}\overline{\cal D}$ amplitudes in terms of tree and penguin
amplitudes.}
\begin{ruledtabular}
\begin{tabular}{rcccrccc}
Mode
          & $T_{1{\cal B}\overline{\cal D}}$
          & $T_{2{\cal B}\overline{\cal D}}$
          & $P_{{\cal B}\overline{\cal D}}$
          & Mode
          & $T^\prime_{1{\cal B}\overline{\cal D}}$
          & $T^\prime_{2{\cal B}\overline{\cal D}}$
          & $P^\prime_{{\cal B}\overline{\cal D}}$
          \\
\hline $B^-\to p \overline{\Delta^{++}}$
          & $-\sqrt6$
          & $\sqrt6$
          & $\sqrt{6}$
          & $B^-\to \Sigma^{+} \overline{\Delta^{++}}$
          & $\sqrt6$
          & $-\sqrt6$
          & $-\sqrt{6}$
          \\
$n \overline{\Delta^+}$
          & $-\sqrt2$
          & $0$
          & ${\sqrt2}$
          & $\Xi^{0} \overline{\Sigma^{*+}}$
          & $\sqrt2$
          & 0
          & $-{\sqrt2}$
          \\
$\Lambda \overline{\Sigma^{*+}}$
          & ${2}/{\sqrt3}$
          & $-{1}/{\sqrt3}$
          & $-{\sqrt3}$
          & $\Lambda \overline{\Delta^+}$
          & $1/\sqrt3$
          & $1/\sqrt3$
          & $0$
          \\
$\Sigma^0 \overline{\Sigma^{*+}}$
          & $0$
          & $-1$
          & $-{1}$
          & $\Sigma^{0} \overline{\Delta^+}$
          & $-1$
          & $1$
          & ${2}$
          \\
$\Sigma^{-} \overline{\Sigma^{*0}}$
          & 0
          & 0
          & $ -{1}$
          & $\Sigma^{-} \overline{\Delta^{0}}$
          & 0
          & 0
          & $ {\sqrt2}$
          \\
$\Xi^{-} \overline{\Xi^{*0}}$
          & $0$
          & 0
          & $-{\sqrt2}$
          & $\Xi^{-} \overline{\Sigma^{*0}}$
          & 0
          & 0
          & ${1}$
          \\
\hline $\overline B {}^0\to p \overline{\Delta^+}$
          & $-\sqrt2$
          & $\sqrt2$
          & ${\sqrt2}$
          & $\overline B {}^0\to \Sigma^{+} \overline{\Delta^{+}}$
          & $\sqrt2$
          & $-\sqrt2$
          & $-{\sqrt2}$
          \\
$n \overline{\Delta^0}$
          & $-\sqrt2$
          & 0
          & ${\sqrt2}$
          & $\Xi^{0} \overline{\Sigma^{*0}}$
          & $1$
          & $0$
          & $-{1}$
          \\
$\Lambda \overline{\Sigma^{*0}}$
          & $\sqrt{{2}/{3}}$
          & $-{1}/{\sqrt6}$
          & $-{\sqrt{3/2}}$
          & $\Lambda \overline{\Delta^0}$
          & ${1}/{\sqrt3}$
          & ${1}/{\sqrt3}$
          & $0$
          \\
$\Sigma^{0} \overline{\Sigma^{*0}}$
          & $0$
          & $-{1}/{\sqrt2}$
          & $-{1}/{\sqrt2}$
          & $\Sigma^{0} \overline{\Delta^0}$
          & $-1$
          & $1$
          & ${2}$
          \\
$\Sigma^{-} \overline{\Sigma^{*-}}$
          & 0
          & 0
          & $-{\sqrt2}$
          & $\Sigma^{-} \overline{\Delta^{-}}$
          & $0$
          & 0
          & $\sqrt{6}$
          \\
$\Xi^{-} \overline{\Xi^{*-}}$
          & 0
          & 0
          & $-{\sqrt2}$
          & $\Xi^{-} \overline{\Sigma^{*-}}$
          & 0
          & 0
          & ${\sqrt2}$
          \\
\hline $\overline B {}^0_s\to p \overline{\Sigma^{*+}}$
          & $-\sqrt2$
          & $\sqrt2$
          & ${\sqrt2}$
          & $\overline B {}^0_s\to \Sigma^{+} \overline{\Sigma^{*+}}$
          & $\sqrt2$
          & $-\sqrt2$
          & $-{\sqrt2}$
          \\
$n \overline{\Sigma^{*0}}$
          & $-1$
          & $0$
          & ${1}$
          & $\Xi^0 \overline{\Xi^{*0}}$
          & ${\sqrt2}$
          & $0$
          & $-{\sqrt2}$
          \\
$\Lambda \overline{\Xi^{*0}}$
          & ${2}/{\sqrt3}$
          & $-{1}/{\sqrt3}$
          & $-{\sqrt3}$
          & $\Lambda \overline{\Sigma^{*0}}$
          & ${1}/{\sqrt6}$
          & ${1}/{\sqrt6}$
          & $0$
          \\
$\Sigma^{0} \overline{\Xi^{*0}}$
          & $0$
          & $-1$
          & $-{1}$
          & $\Sigma^{0} \overline{\Sigma^{*0}}$
          & $-{1}/{\sqrt2}$
          & ${1}/{\sqrt2}$
          & ${\sqrt2}$
          \\
$\Sigma^{-} \overline{\Xi^{*-}}$
          & 0
          & 0
          & $-{\sqrt2}$
          & $\Sigma^{-} \overline{\Sigma^{*-}}$
          & 0
          & 0
          & ${\sqrt2}$
          \\
$\Xi^{-} \overline{\Omega^{-}}$
          & 0
          & 0
          & $-\sqrt{6}$
          & $\Xi^{-} \overline{\Xi^{*-}}$
          & 0
          & 0
          & ${\sqrt2}$
          \\
\end{tabular}
\end{ruledtabular}
\end{table*}

We are now ready to obtain the effective Hamiltonian for the
$B\to{\cal B}\overline{\cal D}$ decays. By replacing
$\overline{\cal D}_{ikl}$ in Eq.~(\ref{eq:DD}) by
$\epsilon_{ika}\overline {\cal B}^a_l$ and
$\epsilon_{akl}\overline {\cal B}^a_i$, we have
\begin{eqnarray}
H_{\rm eff}&=& -\sqrt{6}\,T_{1{\cal B}\overline{\cal D}}\,
                        \overline B_m H^{ik}_j
                        \epsilon_{ika}\overline {\cal B}^a_l
                        {\cal D}^{ljm}
\nonumber\\
           &&          -\sqrt{6}\,T_{2{\cal B}\overline{\cal D}}\,
                        \overline B_m H^{ik}_j
                        \epsilon_{akl}\overline {\cal B}^a_i
                       {\cal D}^{ljm}
\nonumber\\
           && -\sqrt{6} P_{{\cal B}\overline{\cal D}}\,
                        \overline B_m H^k
                        \epsilon_{kia} \overline {\cal B}^a_{l}
                       {\cal D}^{lim},
\label{eq:BD}
\end{eqnarray}
where some pre-factors are introduced for later purpose. As argued
previously, we shall concentrate on $T$ and $P$ terms. Note that
we have two tree and one penguin amplitudes. We use $T^\prime$ and
$P^\prime$ for the $|\Delta S|=1$ case. The resulting
decompositions are shown in Table~\ref{tab:BD}.

\begin{table*}[t!]
\caption{\label{tab:DB} Decomposition of $\overline B\to {\cal
D}\overline{\cal B}$ amplitudes in terms of tree and penguin
amplitudes.}
\begin{ruledtabular}
\begin{tabular}{rcccrccc}
Mode
          & $T_{1{\cal D}\overline{\cal B}}$
          & $T_{2{\cal D}\overline{\cal B}}$
          & $P_{{\cal D}\overline{\cal B}}$
          & Mode
          & $T^\prime_{1{\cal D}\overline{\cal B}}$
          & $T^\prime_{2{\cal D}\overline{\cal B}}$
          & $P^\prime_{{\cal D}\overline{\cal B}}$
          \\
\hline $B^-\to \Delta^0 \overline{p}$
          & $\sqrt2$
          & $0$
          & $-{\sqrt2}$
          & $B^-\to \Sigma^{*0} \overline{p}$
          & $1$
          & $0$
          & $-{1}$
          \\
$\Sigma^{*0} \overline{\Sigma^{+}}$
          & $-1$
          & $0$
          & ${1}$
          & $\Xi^{*0} \overline{\Sigma^{+}}$
          & $ -\sqrt2$
          & $0$
          & ${\sqrt2}$
          \\
$\Sigma^{*-} \overline{\Lambda}$
          & $0$
          & $0$
          & ${\sqrt3}$
          &$\Xi^{*-} \overline{\Lambda}$
          & $0$
          & $0$
          & ${\sqrt3}$
          \\
$\Sigma^{*-} \overline{\Sigma^{0}}$
          & 0
          & $0$
          & $-{1}$
          & $\Xi^{*-} \overline{\Sigma^{0}}$
          & 0
          & $0$
          & $-{1}$
          \\
$\Xi^{*-} \overline{\Xi^{0}}$
          & $0$
          & $0$
          & ${\sqrt2}$
          & $\Omega^{-} \overline{\Xi^{0}}$
          & 0
          & $0$
          & $\sqrt{6}$
          \\
$\Delta^- \overline{n}$
          & $0$
          & $0$
          & $-\sqrt{6}$
          & $\Sigma^{*-} \overline{n}$
          & $0$
          & $0$
          & $-{\sqrt2}$
          \\
\hline $\overline B {}^0\to \Sigma^{*0} \overline{\Sigma^{0}}$
          & ${1}/{\sqrt2}$
          & 0
          & $-{1}/{\sqrt2}$
          & $\overline B {}^0\to \Xi^{*0} \overline{\Sigma^{0}}$
          & $1$
          & 0
          & $-{1}$
          \\
$\Delta^0 \overline{n}$
          & $\sqrt2$
          & $-\sqrt2$
          & ${\sqrt2}$
          & $\Sigma^{*0} \overline{n}$
          & $1$
          & $-1$
          & ${1}$
          \\
$\Sigma^{*0} \overline{\Lambda}$
          & $-{1}/{\sqrt6}$
          & $\sqrt{{2}/{3}}$
          & $-{\sqrt{3/2}}$
          & $\Xi^{*0} \overline{\Lambda}$
          & $-{1}/{\sqrt3}$
          & ${2}/{\sqrt3}$
          & $-{\sqrt3}$
          \\
$\Delta^+ \overline{p}$
          & $0$
          & $-\sqrt2$
          & ${\sqrt2}$
          & $\Sigma^{*+} \overline{p}$
          & 0
          & $-\sqrt2$
          & ${\sqrt2}$
          \\
$\Sigma^{*-} \overline{\Sigma^{-}}$
          & 0
          & 0
          & $-{\sqrt2}$
          & $\Xi^{*-} \overline{\Sigma^{-}}$
          & $0$
          & 0
          & $-{\sqrt2}$
          \\
$\Xi^{*-} \overline{\Xi^{-}}$
          & 0
          & 0
          & $-{\sqrt2}$
          & $\Omega^{-} \overline{\Xi^{-}}$
          & $0$
          & 0
          & $-\sqrt{6}$
          \\

\hline $\overline B {}^0_s\to \Sigma^{*0} \overline{\Xi^{0}}$
          & $-1$
          & 1
          & $-{1}$
          & $\overline B {}^0_s\to \Xi^{*0} \overline{\Xi^{0}}$
          & $-\sqrt2$
          & $\sqrt2$
          & $-{\sqrt2}$
          \\
$\Delta^0 \overline{\Lambda}$
          & $-{2}/{\sqrt3}$
          & ${1}/{\sqrt3}$
          & $0$
          & $\Sigma^{*0} \overline{\Lambda}$
          & $-\sqrt{{2}/{3}}$
          & ${1}/{\sqrt6}$
          & $0$
          \\
$\Delta^0 \overline{\Sigma^{0}}$
          & $0$
          & $-1$
          & ${2}$
          & $\Sigma^{*0} \overline{\Sigma^{0}}$
          & 0
          & $-{1}/{\sqrt2}$
          & ${\sqrt2}$
          \\
$\Delta^+ \overline{\Sigma^{+}}$
          & $0$
          & $\sqrt2$
          & $-{\sqrt2}$
          & $\Sigma^{*+} \overline{\Sigma^{+}}$
          & $0$
          & $\sqrt2$
          & $-{\sqrt2}$
          \\
$\Sigma^{*-} \overline{\Xi^{-}}$
          & $0$
          & $0$
          & ${\sqrt2}$
          & $\Xi^{*-} \overline{\Xi^{-}}$
          & 0
          & $0$
          & ${\sqrt2}$
          \\
$\Delta^{-} \overline{\Sigma^{-}}$
          & $0$
          & $0$
          & $\sqrt{6}$
          & $\Sigma^{*-} \overline{\Sigma^{-}}$
          & $0$
          & $0$
          & ${\sqrt2}$
          \\
\end{tabular}
\end{ruledtabular}
\end{table*}

There are relations on decay amplitudes. By using
Table~\ref{tab:BD}, we have
\begin{widetext}
\begin{eqnarray}
\sqrt2\,A(B^-\to n\overline{\Delta^+})
 &+&\sqrt3\,A(B^-\to \Lambda\overline{\Sigma^{*+}})
 -A(B^-\to \Sigma^0\overline{\Sigma^{*+}})=0,
\nonumber\\
A(B^-\to p\overline{\Delta^{++}})
 &+&\sqrt3\,A(B^-\to n\overline{\Delta^{+}})
 +3\sqrt2\,A(B^-\to \Lambda\overline{\Sigma^{*+}})
 -\sqrt6\,A(B^-\to \Sigma^-\overline{\Sigma^{*0}})=0,
\nonumber\\
 A(B^-\to p\overline{\Delta^{++}})
  &=&\sqrt3 A(\overline B {}^0\to p\overline{\Delta^+})
  =\sqrt3 A(\overline B {}^0_s\to p\overline{\Sigma^{*+}}),
 \nonumber\\
 A(B^-\to n\overline{\Delta^+})
  &=&A(\overline B {}^0\to n\overline{\Delta^0})
  =\sqrt2 A(\overline B {}^0_s\to n\overline{\Sigma^{*+}}),
 \nonumber\\
 A(B^-\to\Lambda(\Sigma^0)\overline{\Sigma^{*+}})
  &=&\sqrt2 A(\overline B {}^0\to\Lambda(\Sigma^0)\overline{\Sigma^{*0}})
  =A(\overline B
  {}^0_s\to\Lambda(\Sigma^0)\overline{\Xi^{*0}}),
  \nonumber\\
 \sqrt6 A(B^-\to\Sigma^-\overline{\Sigma^{*0}})
  &=&\sqrt3 A(\overline B {}^0\to \Sigma^-\overline{\Sigma^{*-}})
  =\sqrt3 A(\overline B {}^0_s\to \Sigma^-\overline{\Xi^{*-}})
 \nonumber\\
 &=&\sqrt3 A(B^-\to\Xi^-\overline{\Xi^{*0}})
  =\sqrt3 A(\overline B {}^0\to \Xi^-\overline{\Xi^{*-}})
  =A(\overline B {}^0_s\to \Xi^-\overline{\Omega^{-}}),
  \label{eq:relBD0}
 \end{eqnarray}
for the $\Delta S=0$ case and
 \begin{eqnarray}
 \sqrt2\,A(B^-\to \Xi^0\overline{\Sigma^{*+}})
   &-&\sqrt3\,A(B^-\to\Lambda\overline{\Delta^{+}})
   +A(B^-\to \Sigma^0\overline{\Delta^{+}})
   =0,
 \nonumber\\
 A(B^-\to \Sigma^+\overline{\Delta^{++}})
   &-&2\sqrt3\,A(B^-\to \Xi^0\overline{\Sigma^{*+}})
   +3\sqrt2\,A(B^-\to\Lambda\overline{\Delta^{+}})
   -\sqrt3\,A(B^-\to\Sigma^-\overline{\Delta^{0}})=0,
 \nonumber\\
  A(B^-\to \Sigma^+\overline{\Delta^{++}})
  &=&\sqrt3 A(\overline B {}^0\to \Sigma^+\overline{\Delta^+})
  =\sqrt3 A(\overline B {}^0_s\to \Sigma^+\overline{\Sigma^{*+}}),
 \nonumber\\
 A(B^-\to \Xi^0\overline{\Sigma^{*+}})
  &=&\sqrt2 A(\overline B {}^0\to \Xi^0\overline{\Sigma^{*0}})
  =A(\overline B {}^0_s\to \Xi^0\overline{\Xi^{*0}}),
 \nonumber\\
 A(B^-\to\Lambda(\Sigma^0)\overline{\Delta^{+}})
  &=&A(\overline B {}^0\to\Lambda(\Sigma^0)\overline{\Delta^{0}})
  =\sqrt2 A(\overline B
  {}^0_s\to\Lambda(\Sigma^0)\overline{\Sigma^{*0}}),
 \nonumber\\
 \sqrt3 A(B^-\to\Sigma^-\overline{\Delta^{0}})
  &=&A(\overline B {}^0\to \Sigma^-\overline{\Delta^{-}})
  =\sqrt3 A(\overline B {}^0_s\to \Sigma^-\overline{\Sigma^{*-}})
 \nonumber\\
  &=&\sqrt6 A(B^-\to\Xi^-\overline{\Sigma^{*0}})
  =\sqrt3 A(\overline B {}^0\to \Xi^-\overline{\Sigma^{*-}})
  =\sqrt3 A(\overline B {}^0_s\to \Xi^-\overline{\Xi^{*-}}),
  \label{eq:relBD1}
 \end{eqnarray}
 \end{widetext}
 for the $|\Delta S|=1$ case.

In Table~\ref{tab:BD} we find that the penguin-dominated ($|\Delta
S|=1$) $\overline B \to \Lambda \overline {\cal D}$ decays do not
receive any penguin contribution. This can be easily understood.
In the penguin topology, the $s$ quark of $\Lambda$ is from the
$b\to s$ decay, so the $u$, $d$ quarks of $\Lambda$ are correlated
to their pair-creating partners, the $\bar u$, $\bar d$
anti-quarks, in the accompanying anti-decuplet baryon. The
anti-symmetry property of the $\Lambda$ wave function and the
symmetry property of the anti-decuplet wave function are
responsible for the vanishing of the corresponding penguin
amplitudes.

We now turn to the $B\to{\cal D}\overline{\cal B}$ case. By
replacing ${\cal D}^{ljm}$ in Eq.~(\ref{eq:DD}) by $\epsilon^{ljb}
{\cal B}^m_b$ and $\epsilon^{bjm} {\cal B}^l_b$ for the decays, we
have
\begin{eqnarray}
H_{\rm eff}&=& -\sqrt{6}\,T_{1{\cal D} \overline{\cal
                             B}}\,
                        \overline B_m H^{ik}_j \overline{\cal D}_{ikl}
                        \epsilon^{ljb} {\cal B}^m_b
 \nonumber\\
           &&              -\sqrt{6}\,T_{2{\cal D} \overline{\cal
                             B}}\,
                        \overline B_m H^{ik}_j \overline{\cal D}_{ikl}
                        \epsilon^{bjm} {\cal B}^l_b
\nonumber\\
           && +\sqrt{6}\, P_{{\cal D}\overline{\cal B}}\,
             \overline B_m H^k \overline{\cal D}_{kil} \epsilon^{bim}
             {\cal B}^l_{b}.
\label{eq:DB}
\end{eqnarray}
We have two tree and one penguin amplitudes.

The decomposition of decay amplitudes are shown in
Table~\ref{tab:DB}. From the table, we have
\begin{widetext}
\begin{eqnarray}
\sqrt3\,A(B^-\to\Delta^0\overline p)
 &+&\sqrt3\,A(\overline B {}^0\to\Delta^+\overline p)
 =A(B^-\to\Delta^-\overline n)
 +\sqrt3\,A(\overline B {}^0\to\Delta^0\overline n),
 \nonumber\\
 A(B^-\to\Delta^0\overline p)
 &=&-\sqrt2\,A(B^-\to\Sigma^{*0}\overline{\Sigma^+})
 =2\,A(\overline B {}^0\to\Sigma^{*0}\overline{\Sigma^0}),
\nonumber\\
A(\overline B {}^0\to\Delta^0\overline n)
 &=&-\sqrt2\,A(\overline B {}^0_s\to\Sigma^{*0}\overline{\Xi^0}),
\nonumber\\
 A(\overline B {}^0\to\Delta^+\overline p)
 &=&-A(\overline B {}^0_s\to\Delta^+\overline{\Sigma^+}),
 \nonumber\\
A(B^-\to\Delta^-\overline n)
 &=&-\sqrt2\,A(B^-\to\Sigma^{*-}\overline\Lambda)
 =\sqrt6\,A(B^-\to\Sigma^{*-}\overline{\Sigma^0})
\nonumber\\
 &=&-\sqrt3\,A(B^-\to\Xi^{*-}\overline{\Xi^0})
 =\sqrt3\, A(\overline B {}^0\to\Xi^{*-}\overline{\Xi^-})
 =\sqrt3\, A(\overline B
{}^0\to\Sigma^{*-}\overline{\Sigma^-})
\nonumber\\
&=&-\sqrt3\, A(\overline B {}^0_s\to\Sigma^{*-}\overline{\Xi^-})
=-A(\overline B {}^0_s\to\Delta^{-}\overline{\Sigma^-}),
\label{eq:relDB0}
\end{eqnarray}
for $\Delta S=0$ processes, and
\begin{eqnarray}
\sqrt2\,A(B^-\to\Sigma^{*0}\overline p)
 &+&A(\overline B {}^0\to\Sigma^{*+}\overline p)
 =A(B^-\to\Sigma^{*-}\overline n)
 +\sqrt2\,A(\overline B {}^0\to\Sigma^{*0}\overline n),
 \nonumber\\
\sqrt2\,A(B^-\to\Sigma^{*0}\overline p)
 &=&-A(B^-\to\Xi^{*0}\overline{\Sigma^+})
 =\sqrt2\,A(\overline B {}^0\to\Xi^{*0}\overline{\Sigma^0}),
\nonumber\\
\sqrt2\,A(\overline B {}^0\to\Sigma^{*0}\overline n)
 &=&-A(\overline B {}^0_s\to\Xi^{*0}\overline{\Xi^0}),
\nonumber\\
 A(\overline B {}^0\to\Sigma^{*+}\overline p)
 &=&-A(\overline B {}^0_s\to\Sigma^{*+}\overline{\Sigma^+}),
 \nonumber\\
\sqrt3\,A(B^-\to\Sigma^{*-}\overline n)
 &=&-A(B^-\to\Omega^{-}\overline{\Xi^0})
 =A(\overline B {}^0\to\Omega^{-}\overline{\Xi^-})
 \nonumber\\
&=&-\sqrt2\,A(B^-\to\Xi^{*-}\overline\Lambda)
 =\sqrt6\,A(B^-\to\Xi^{*-}\overline{\Sigma^0})
 =\sqrt3\, A(\overline B {}^0\to\Xi^{*-}\overline{\Sigma^-})
\nonumber\\
 &=&-\sqrt3\,A(\overline B{}^0_s\to\Xi^{*-}\overline{\Xi^-})
 =-\sqrt3\,A(\overline B {}^0_s\to\Sigma^{*-}\overline{\Sigma^-}),
 \label{eq:relDB1}
\end{eqnarray}
\end{widetext}
for $|\Delta S|=1$ processes. The implication of
Table~\ref{tab:BD} and Table~\ref{tab:DB} on the phenomenology of
the corresponding decay modes will be discussed later.

\subsection{\boldmath $\overline B$ to octet anti-octet baryonic decays}

\begin{table*}[t!]
\caption{\label{tab:BB0} Decomposition of $\overline B\to {\cal
B}\overline{\cal B}$ amplitudes for $\Delta S=0$ transitions in
terms of tree and penguin amplitudes.}
\begin{ruledtabular}
\begin{tabular}{rcccccc}
Mode
          & $T_{1{\cal B}\overline{\cal B}}$
          & $T_{2{\cal B}\overline{\cal B}}$
          & $T_{3{\cal B}\overline{\cal B}}$
          & $T_{4{\cal B}\overline{\cal B}}$
          & $P_{1{\cal B}\overline{\cal B}}$
          & $P_{2{\cal B}\overline{\cal B}}$
          \\
\hline $B^-\to n \overline{p}$
          & $-1$
          & 0
          & $0$
          & 0
          & $-{5}$
          & 0
          \\
$\Lambda\overline{\Sigma^+}$
          & $-\sqrt{{2}/{3}}$
          & 0
          & ${1}/{\sqrt6}$
          & 0
          & $-{5}/{\sqrt6}$
          & $-{5}/{\sqrt6}$
          \\
$\Sigma^{-} \overline{\Lambda}$
          & $0$
          & 0
          & 0
          & 0
          & $-{5}/{\sqrt6}$
          & $-{5}/{\sqrt6}$
          \\
$\Sigma^{-} \overline{\Sigma^0}$
          & $0$
          & 0
          & 0
          & 0
          & $-{5}/{\sqrt2}$
          & ${5}/{\sqrt2}$
          \\
$\Sigma^0 \overline{\Sigma^+}$
          & 0
          & 0
          & ${1}/{\sqrt2}$
          & 0
          & ${5}/{\sqrt2}$
          & $-{5}/{\sqrt2}$
          \\
$\Xi^{-} \overline{\Xi^0}$
          & $0$
          & 0
          & 0
          & 0
          & 0
          & $-{5}$
          \\
\hline $\overline B {}^0\to p \overline{p}$
          & 0
          & 1
          & 0
          & $-1$
          & 0
          & ${5}$
          \\
 $n \overline{n}$
          & $-1$
          & 1
          &0
          &0
          & $-{5}$
          & ${5}$
          \\
$\Lambda \overline{\Lambda}$
          & $-{1}/{3}$
          & ${2}/{3}$
          & ${1}/{6}$
          & $-{1}/{3}$
          & $-{5}/{6}$
          & ${25}/{6}$
          \\
$\Lambda \overline{\Sigma^0}$
          & ${1}/{\sqrt3}$
          & 0
          & $-{1}/{2\sqrt3}$
          & $0$
          & ${5}/{2\sqrt3}$
          & ${5}/{2\sqrt3}$
          \\
$\Sigma^{0} \overline{\Lambda}$
          & 0
          & 0
          & ${1}/{2\sqrt3}$
          & $-{1}/{\sqrt3}$
          & ${5}/{2\sqrt3}$
          & ${5}/{2\sqrt3}$
          \\
$\Sigma^{-} \overline{\Sigma^{-}}$
          & $0$
          & 0
          & 0
          & 0
          & $-{5}$
          & ${5}$
          \\
$\Sigma^{0} \overline{\Sigma^0}$
          & 0
          & 0
          & $-{1}/{2}$
          & 0
          & $-{5}/{2}$
          & ${5}/{2}$
          \\
$\Xi^{-} \overline{\Xi^{-}}$
          & $0$
          & 0
          & 0
          & 0
          & 0
          & ${5}$
          \\
\hline $\overline B {}^0_s\to p \overline{\Sigma^{+}}$
          & 0
          & $-1$
          & 0
          & 1
          & 0
          & $-{5}$
          \\
$n \overline{\Lambda}$
          & $\sqrt{{2}/{3}}$
          & $-{1}/{\sqrt6}$
          & 0
          & 0
          & ${10}/{\sqrt6}$
          & $-{5}/{\sqrt6}$
          \\
$n \overline{\Sigma^{0}}$
          & 0
          & ${1}/{\sqrt2}$
          & 0
          & 0
          & 0
          & ${5}/{\sqrt2}$
          \\
$\Lambda \overline{\Xi^{0}}$
          & $-\sqrt{{2}/{3}}$
          & $\sqrt{{2}/{3}}$
          & ${1}/{\sqrt6}$
          & $-{1}/{\sqrt6}$
          & $-{5}/{\sqrt6}$
          & ${10}/{\sqrt6}$
          \\
$\Sigma^0 \overline{\Xi^0}$
          & 0
          & 0
          & ${1}/{\sqrt2}$
          & $-{1}/{\sqrt2}$
          & ${5}/{\sqrt2}$
          & 0
          \\
$\Sigma^{-} \overline{\Xi^-}$
          & $0$
          & 0
          & 0
          & 0
          & $-{5}$
          & 0
\end{tabular}
\end{ruledtabular}
\end{table*}

The decomposition of $\overline B\to {\cal B}\overline {\cal B}$
amplitudes can be achieved similarly. To obtain the corresponding
decay effective Hamiltonian, we replace $\overline{\cal D}_{ikl}$
(${\cal D}^{ljm}$) in Eq.~(\ref{eq:DD}) by
$\epsilon_{ika}\overline {\cal B}^a_l$ and
$\epsilon_{akl}\overline {\cal B}^a_i$ ($\epsilon^{ljb} {\cal
B}^m_b$ and $\epsilon^{bjm} {\cal B}^l_b$).
We have
\begin{eqnarray}
H_{\rm eff}&=& T_{1{\cal B} \overline{\cal B}}\,
                        \overline B_m H^{ik}_j
                        \epsilon_{ika}\overline {\cal B}^a_l
                        \epsilon^{ljb} {\cal B}^m_b
\nonumber\\&&                         +T_{2{\cal B} \overline{\cal
B}}\,
                        \overline B_m H^{ik}_j
                        \epsilon_{ika}\overline {\cal B}^a_l
                        \epsilon^{bjm} {\cal B}^l_b
\nonumber\\&&                       +T_{3{\cal B} \overline{\cal
B}}\,
                        \overline B_m H^{ik}_j
                        \epsilon_{akl}\overline {\cal B}^a_i
                        \epsilon^{ljb} {\cal B}^m_b
\nonumber\\&&
                           +T_{4{\cal B} \overline{\cal B}}\,
                        \overline B_m H^{ik}_j
                        \epsilon_{akl}\overline {\cal B}^a_i
                        \epsilon^{bjm} {\cal B}^l_b
\nonumber\\&&         -{5}\, P_{1{\cal B}\overline{\cal B}}\,
                        \overline B_m H^k
                        \epsilon_{kia} \overline {\cal B}^a_{l}
                        \epsilon^{lib} {\cal B}^m_{b}
\nonumber\\&&         -{5}\, P_{2{\cal B}\overline{\cal B}}\,
                        \overline B_m H^k
                        \epsilon_{kia} \overline {\cal B}^a_{l}
                        \epsilon^{bim} {\cal B}^l_{b}.
\label{eq:BB}
\end{eqnarray}
The coefficients are assigned for later purpose.

There are four tree amplitudes and two penguin amplitudes. For the
tree amplitudes all four combinations as suggested in
Eqs.~(\ref{eq:qqq}, \ref{eq:identity}) are used. For the penguin
part, only two of the four combinations are independent. The two
combinations used in the penguin amplitudes can be expressed as
\begin{eqnarray}
 \epsilon_{kia} \overline {\cal B}^a_{l}
 \epsilon^{lib} {\cal B}^m_{b}
&=&\frac{1}{2}\{{\cal B},\overline{\cal B}\}^m_k
         +\frac{1}{2}[{\cal B},\overline{\cal B}]^m_k,
\nonumber\\
 \epsilon_{kia} \overline {\cal B}^a_{l}
 \epsilon^{bim} {\cal B}^l_{b}
&=&\frac{1}{2}\{{\cal B},\overline{\cal B}\}^m_k
         -\frac{1}{2}[{\cal B},\overline{\cal B}]^m_k
         -\delta^m_k {\rm Tr}({\cal B}\overline{\cal B}).
\nonumber\\
\end{eqnarray}
One can readily recognize the anti-commutator and the commutator
parts as $D$ and $F$-terms in the usual SU(3) combination when
dealing with an operator having $ \bar q_k H^k_m q^m$ flavor
quantum number~\cite{text}. The ${\rm Tr}({\cal B}\overline{\cal
B})$ term, as denoted as $S$ in Ref.~\cite{Chua:2002wn}, is needed
in a non-traceless $H^k_m$ case. If we use the $D$-$F$-$S$ basis
instead of the $\epsilon\,\overline{\cal B} \,\epsilon\,{\cal B}$
basis, we will have
\begin{equation}
P_{D}=\frac{1}{2}(P_1+P_2),\quad P_F=\frac{1}{2}(P_1-P_2),\quad
P_S=-P_2. \label{eq:DFS}
\end{equation}
Since we only have two independent components, we have a
constraint: $P_D-P_F+P_S=0$. This is nothing but the
Okubo-Zweig-Iizuka (OZI) rule in the ${\cal B}\overline {\cal B}$
sector. A similar constraint is used in Ref.~\cite{Chua:2002wn},
where it is enforced by handed. The OZI rule is satisfied
automatically in this approach, since we are matching the baryon
quark flavors (see, Fig.~\ref{fig:TPE}) in the beginning.

\begin{table*}[t!]
\caption{\label{tab:BB1} Decomposition of $\overline B\to {\cal
B}\overline{\cal B}$ amplitudes for $|\Delta S|=1$ transitions in
terms of tree and penguin amplitudes.}
\begin{ruledtabular}
\begin{tabular}{rcccccc}
Mode
          & $T^\prime_{1{\cal B}\overline{\cal B}}$
          & $T^\prime_{2{\cal B}\overline{\cal B}}$
          & $T^\prime_{3{\cal B}\overline{\cal B}}$
          & $T^\prime_{4{\cal B}\overline{\cal B}}$
          & $P^\prime_{1{\cal B}\overline{\cal B}}$
          & $P^\prime_{2{\cal B}\overline{\cal B}}$
          \\
\hline $B^-\to \Lambda \overline{p}$
          & ${1}/{\sqrt6}$
          & 0
          & ${1}/{\sqrt6}$
          & 0
          & ${10}/{\sqrt6}$
          & $-{5}/{\sqrt6}$
          \\
$\Xi^0 \overline{\Sigma^+}$
          & $-1$
          & 0
          & 0
          & 0
          & $-{5}$
          & 0
          \\
$\Xi^{-} \overline{\Sigma^0}$
          & $0$
          & 0
          & 0
          & 0
          & $-{5}/{\sqrt2}$
          & 0
          \\
$\Xi^{-} \overline{\Lambda}$
          & $0$
          & 0
          & 0
          & 0
          & $-{5}/{\sqrt6}$
          & ${10}/{\sqrt6}$
          \\
$\Sigma^{-} \overline{n}$
          & $0$
          & 0
          & 0
          & 0
          & 0
          & $-{5}$
          \\
$\Sigma^0\overline{p}$
          & $-{1}/{\sqrt2}$
          & 0
          & ${1}/{\sqrt2}$
          & 0
          & 0
          & $-{5}/{\sqrt2}$
          \\
\hline $\overline B {}^0\to \Lambda \overline{n}$
          & ${1}/{\sqrt6}$
          & $-{1}/{\sqrt6}$
          & ${1}/{\sqrt6}$
          & $-{1}/{\sqrt6}$
          & ${10}/{\sqrt6}$
          & $-{5}/{\sqrt6}$
          \\
$\Xi^- \overline{\Sigma^-}$
          & 0
          & 0
          & 0
          & 0
          & $-{5}$
          & 0
          \\
$\Xi^0 \overline{\Sigma^0}$
          & ${1}/{\sqrt2}$
          & 0
          & 0
          & 0
          & ${5}/{\sqrt2}$
          & 0
          \\
$\Xi^{0} \overline{\Lambda}$
          & $-{1}/{\sqrt6}$
          & $\sqrt{{2}/{3}}$
          & 0
          & 0
          & $-{5}/{\sqrt6}$
          & ${10}/{\sqrt6}$
          \\
$\Sigma^{+}\overline{p}$
          & 0
          & $-1$
          & 0
          & 1
          & 0
          & $-{5}$
          \\
$\Sigma^{0} \overline{n}$
          & $-{1}/{\sqrt2}$
          & ${1}/{\sqrt2}$
          & ${1}/{\sqrt2}$
          & $-{1}/{\sqrt2}$
          & 0
          & ${5}/{\sqrt2}$
          \\
\hline $\overline B {}^0_s\to \Lambda \overline{\Lambda}$
          & $-{1}/{3}$
          & ${1}/{6}$
          & $-{1}/{3}$
          & ${1}/{6}$
          & $-{10}/{3}$
          & ${5}/{3}$
          \\
$\Xi^{0} \overline{\Xi^0}$
          & $-1$
          & 1
          & 0
          & 0
          & $-{5}$
          & ${5}$
          \\
$\Xi^{-} \overline{\Xi^{-}}$
          & $0$
          & 0
          & 0
          & 0
          & $-{5}$
          & ${5}$
          \\
$\Sigma^+ \overline{\Sigma^+}$
          & $0$
          & 1
          & 0
          & $-1$
          & 0
          & ${5}$
          \\
$\Sigma^{0} \overline{\Sigma^0}$
          & 0
          & ${1}/{2}$
          & 0
          & $-{1}/{2}$
          & 0
          & ${5}$
          \\

$\Sigma^{-} \overline{\Sigma^{-}}$
          & $0$
          & 0
          & 0
          & 0
          & 0
          & ${5}$
          \\
$\Lambda \overline{\Sigma^0}$
          & 0
          & $-{1}/{2\sqrt3}$
          & 0
          & $-{1}/{2\sqrt3}$
          & 0
          & $0$
          \\
$\Sigma^0 \overline{\Lambda}$
          & ${1}/{\sqrt3}$
          & $-{1}/{2\sqrt3}$
          & $-{1}/{\sqrt3}$
          & ${1}/{2\sqrt3}$
          & 0
          & $0$
          \\
\end{tabular}
\end{ruledtabular}
\end{table*}

We shown in Table~\ref{tab:BB0}~(\ref{tab:BB1}) the decay
amplitudes for $|\Delta S|=0\,(1)$ processes. From the tables, we
obtain
\begin{eqnarray}
 \sqrt3\,A(B^-\to\Lambda\overline{\Sigma^+})
   &=&A(B^-\to\Sigma^0\overline{\Sigma^+})
\nonumber\\
   &&+\sqrt2 A(B^-\to n\overline{p}),
 \nonumber\\
 \sqrt3\,A(B^-\to\Sigma^-\overline{\Lambda})
   &=&A(B^-\to\Sigma^-\overline{\Sigma^0})
\nonumber\\
   &&+\sqrt2 A(B^-\to \Xi^-\overline{\Xi^0}),
 \nonumber\\
 A(\overline B {}^0\to p\overline p)
 &=&-A(\overline B {}^0_s\to p\overline{\Sigma^+}),
 \nonumber\\
  A(B^-\to \Sigma^0\overline{\Sigma^+})
 &=&-\sqrt2\,A(\overline B {}^0\to \Sigma^0\overline{\Sigma^0}),
 \nonumber\\
A(B^-\to \Xi^-\overline{\Xi^0})
 &=&-A(\overline B {}^0\to \Xi^-\overline{\Xi^-}),
 \label{eq:relBB0}
\end{eqnarray}
for the $\Delta S=0$ case, and
\begin{eqnarray}
\sqrt3\,A(B^-\to\Lambda\overline{p})
   &=&-\sqrt2\,A(B^-\to\Xi^0\overline{\Sigma^+})
\nonumber\\
   &&+ A(B^-\to \Sigma^0\overline{p}),
\nonumber\\
A(B^-\to\Xi^-\overline{\Sigma^0})
   &=&\sqrt3\,A(B^-\to\Xi^-\overline{\Lambda})
   \nonumber\\
   &&+\sqrt2\, A(B^-\to \Sigma^-\overline{n}),
\nonumber\\
A(\overline B {}^0\to\Sigma^+\overline p)
   &=&-A(\overline B {}^0_s\to\Sigma^+\overline {\Sigma^+}),
  \label{eq:relBB1}
 \end{eqnarray}
for the $|\Delta S|=1$ case. More relations in the $|\Delta S|=1$
case can be obtained by neglecting the sub-leading tree
contribution. We have
\begin{widetext}
\begin{eqnarray}
A(B^-\to\Xi^-\overline\Lambda)
 &=&A(\overline B {}^0\to\Xi^0\overline\Lambda),
 \nonumber\\
  A(\overline B {}^0_s\to\Xi^-\overline{\Xi^-})
 &=&A(\overline B {}^0_s\to\Xi^0\overline{\Xi^0}),
\nonumber\\
\sqrt2\,A(B^-\to\Lambda\overline p)
  &=&\sqrt2\,A(\overline B {}^0\to\Lambda\overline n)
  =\sqrt3\,A(\overline B {}^0_s\to\Lambda\overline\Lambda),
 \nonumber\\
 A(B^-\to\Xi^0\overline{\Sigma^+})
  &=&\sqrt2\,A(B^-\to\Xi^-\overline{\Sigma^0})
  =A(\overline B {}^0\to\Xi^-\overline{\Sigma^-})
 =-\sqrt2\,A(\overline B {}^0 \to\Xi^0\overline{\Sigma^0}),
 \nonumber\\
 A(B^-\to\Sigma^-\overline n)
 &=&\sqrt2\,A(B^-\to\Sigma^0\overline p)
 =A(\overline B {}^0\to\Sigma^+\overline p)
 =-\sqrt2\,A(\overline B {}^0\to\Sigma^0\overline n)
 \nonumber\\
 &=&A(\overline B {}^0_s\to\Sigma^+\overline{\Sigma^+})
  =A(\overline B {}^0_s\to\Sigma^0\overline{\Sigma^0})
  =A(\overline B {}^0_s\to\Sigma^-\overline{\Sigma^-}). \label{eq:relBB2}
\end{eqnarray}
\end{widetext}
Rate relations implied by Eq.~(\ref{eq:relBB2}) are consistent
with those obtained in Ref.~\cite{Sheikholeslami:fa}. In fact,
Ref.~\cite{Sheikholeslami:fa} uses a generic SU(3) analysis and
has three independent penguin amplitudes. One can reduce these
amplitudes into two penguin amplitudes by imposing the OZI rule.
As noted before the quark diagram approach is consistent with the
OZI rule (see Eq.~(\ref{eq:DFS})). As a result, we have more
relations.

To summarize, based solely on the flavor structure of $H_{\rm W}$,
we obtain one tree and one penguin amplitudes for the ${\cal
D}\overline{\cal D}$ case; two tree and one penguin amplitudes for
each of the ${\cal B}\overline {\cal D}$ and the ${\cal
B}\overline {\cal D}$ case; four tree and two penguin amplitudes
for the ${\cal B}\overline {\cal B}$ case. In principle, these
amplitudes can be extracted from data. Various relations on decay
rates as implied in Eqs.~(\ref{eq:relDD0}, \ref{eq:relDD1},
\ref{eq:relBD0}, \ref{eq:relBD1}, \ref{eq:relDB0},
\ref{eq:relDB1}, \ref{eq:relBB0}, \ref{eq:relBB1},
\ref{eq:relBB2}) are obtained and can be checked experimentally.
It should be pointed out that although we concentrate in tree and
penguin amplitudes, the extension to include other contributions,
such as $W$-exchange, electroweak penguin, weak annihilation and
penguin annihilation amplitudes, is straightforward.

\section{\label{sec:reduction}Reduction of topological amplitudes}

Up to now we only make use of the flavor structure of the weak
Hamiltonian. Amplitudes are decomposed topologically. In most
cases, we need more than one tree and more than one penguin
amplitudes. Although some relations are obtained, it is a long
term experimental project to verify them. In fact, by taking into
account of the chirality structure of $H_{\rm W}$, the number of
independent amplitudes is reduced in the large $m_B$ limit. It
leads to more predictive results.

In general the charmless two-body decay amplitudes can be
expressed as~\cite{Jarfi:1990ej,Cheng:2001tr}
\begin{eqnarray}
{\cal A}(\overline B\to {\cal B}_1 \overline {\cal B}_2)&=&\bar
u_1(A_{{\cal B}\overline {\cal B}}+\gamma_5 B_{{\cal B}\overline
{\cal B}}) v_2,
\nonumber\\
{\cal A}(\overline B\to {\cal D}_1 \overline {\cal B}_2)&=&i
\frac{q^\mu}{m_B} \bar u^\mu_1(A_{{\cal D}\overline {\cal
B}}+\gamma_5 B_{{\cal D}\overline {\cal B}}) v_2,
\nonumber\\
{\cal A}(\overline B\to {\cal B}_1 \overline {\cal D}_2)&=&i
\frac{q^\mu}{m_B}\bar u_1(A_{{\cal B}\overline {\cal D}}+\gamma_5
B_{{\cal B}\overline {\cal D}}) v^\mu_2,
\nonumber\\
{\cal A}(\overline B\to {\cal D}_1 \overline {\cal D}_2)&=&\bar
u^\mu_1(A_{{\cal D}\overline {\cal D}}+\gamma_5 B_{{\cal
D}\overline {\cal D}}) v_{2\mu}
\nonumber\\
 && +\frac{q^\mu q^\nu}{m^2_B}\bar
u^\mu_1(C_{{\cal D}\overline {\cal D}}+\gamma_5 D_{{\cal
D}\overline {\cal D}}) v_{2\nu},
\end{eqnarray}
where $q=p_1-p_2$ and $u^\mu,\,v^\mu$ are the Rarita-Schwinger
vector spinors for a spin-$\frac{3}{2}$ particle. The vector
spinors in various helicity states can be expressed
as~\cite{Moroi:1995fs}
 $u_\mu(\pm\frac{3}{2})=\epsilon_\mu(\pm1)
 u(\pm\frac{1}{2})$,
 $u_\mu(\pm\frac{1}{2})=
 (\epsilon_\mu(\pm1)
 u(\mp\frac{1}{2})
 +\sqrt{2}\,\epsilon_\mu(0)
u(\pm\frac{1}{2}))/\sqrt3$,
 where $\epsilon_\mu(\lambda)$ and
$u(s)$ are the usual polarization vector and spinor, respectively.
By using
$q\cdot\epsilon(\lambda)_{1,2}=\mp\,\delta_{\lambda,0}\,m_B\,
p_c/m_{1,2}$, where $p_c$ is the baryon momentum in the $B$ rest
frame, and the fact that
$\epsilon^*_1(0)\cdot\epsilon_2(0)=(m_B^2-m^2_1-m^2_2)/2m_1 m_2$
is the dominant term among $\epsilon^*_1(\lambda_1)\cdot
\epsilon_2(\lambda_2)$, we have
\begin{eqnarray}
{\cal A}(\overline B\to {\cal D}_1 \overline {\cal B}_2)&=&-i
\sqrt{\frac{2}{3}}\frac{p_c}{m_1} \bar u_1(A_{{\cal D}\overline
{\cal B}}+\gamma_5 B_{{\cal D}\overline {\cal B}}) v_2,
\nonumber\\
{\cal A}(\overline B\to {\cal B}_1 \overline {\cal D}_2)&=&i
\sqrt{\frac{2}{3}}\frac{p_c}{m_2}\bar u_1(A_{{\cal B}\overline
{\cal D}}+\gamma_5 B_{{\cal B}\overline {\cal D}}) v_2,
\nonumber\\
{\cal A}(\overline B\to {\cal D}_1 \overline {\cal
D}_2)&\simeq&\frac{m_B^2}{3m_1m_2}\bar u_1(A^\prime_{{\cal
D}\overline {\cal D}} +\gamma_5 B^\prime_{{\cal D}\overline {\cal
D}})v_2, \label{eq:largemB}
\end{eqnarray}
where $A^\prime_{{\cal D}\overline {\cal D}}=A_{{\cal D}\overline
{\cal D}}-2(p_c/m_B)^2 C_{{\cal D}\overline {\cal D}}$,
$B^\prime_{{\cal D}\overline {\cal D}}=B_{{\cal D}\overline {\cal
D}}-2(p_c/m_B)^2 D_{{\cal D}\overline {\cal D}}$ and decuplets are
in $h=\pm\frac{1}{2}$ helicity states. Note that the first two
equations are exact while the last one holds in the $m_B^2\gg
m^2_{1,2},\,m_1m_2$ limit.
%
For these amplitudes to have the same $m_B^2=(p_1+p_2)^2$ behavior
in the large $m^2_B$ limit, the extra factor of $p_c$ and $m_B$ in
Eq.~(\ref{eq:largemB}) should be suitably compensated by the
$1/m_B^2$ power of the corresponding $A$ and $B$ terms. For all
four $\overline B\to {\mathbf B}_1\overline {\mathbf B}_2$
(${\mathbf B}\overline {\mathbf B}={\cal B} \overline {\cal B}$,
${\cal D} \overline {\cal B}$, ${\cal B} \overline {\cal D}$,
${\cal D} \overline {\cal D}$) decays, we can effectively use
\begin{equation}
{\cal A}(\overline B\to {\mathbf B}_1\overline {\mathbf B}_2)
 =\bar u_1(A+\gamma_5 B)v_2,
 \label{eq:asymptoticform}
\end{equation}
as the decay amplitudes.

The chiral structure of weak interaction provide further
information on $A$ and $B$. For example, in the $\Delta S=0$
processes, we have $b\to u_L\bar u_R d_L$ and $b\to d_L
q_{L(R)}\bar q_{R(L)} $ decays. The produced $d_L$ quark is
left-handed in both tree and penguin decays. Furthermore, as
strong interaction is chirality conserving, the pop up quark pair
$q^\prime\bar q^\prime$ should be $q^\prime_{L}\bar q^\prime_{R}$
or $q^\prime_{R}\bar q^\prime_{L}$. From the conservation of
helicity, the produced baryon and anti-baryon must be in
left-helicity states. To be more specify, we take the $\overline B
{}^0_s\to {\mathbf B}\overline {\mathbf B}$ decay as an example.
According to the above argument, the final state (anti-)baryons
should have $u_L d_L q^\prime_R$ ($\bar q^\prime_L\bar u_R \bar
s_L$) configuration from the tree $b\to u_L\bar u_R d_L$ decay and
$d_L q_{L(R)} q^\prime {}_{R(L)}$ ($\bar q^\prime {}_{L(R)} \bar
q_{R(L)} \bar s_L$) configuration from the penguin $b\to d_L
q_{L(R)}\bar q_{R(L)}$ decay. In the large $m_B$ limit, as the
spinor helicity identify to chirality, we should have $B\to -A$ in
the above equation.

The number of independent topological amplitudes is reduced in the
large $m_B$ limit. The corresponding asymptotic relations are
derived in the appendix. Some of the reduction can be understood
by using the chirality structure of weak interaction, the helicity
argument and the anti-symmetry property of the $\Lambda$ wave
function.

The $b\to u_L \bar u_R d_L$ process have final state $u$ and $d$
quarks both with left-handed chirality or helicity (as light quark
masses are neglected). It is well known that in the $\Lambda$ wave
function the $u$ and $d$ quarks have opposite helicity (as shown
in Eq.~(\ref{eq:wavefunction})). Therefore, the tree amplitudes in
the $\Delta S=0$, $\overline B \to \Lambda \overline {\cal
D},\,\Lambda\overline{\cal B}$ decays should be vanishing. As
shown in Table~\ref{tab:BD} and Table~\ref{tab:BB0} these
amplitudes are proportional to $2\,T_{1{\cal B}\overline{\cal
D}}-T_{2{\cal B}\overline{\cal D}}$ and $2 T_{1{\cal B}
\overline{\cal B}} - T_{3{\cal B} \overline{\cal B}}$, $2
T_{2{\cal B} \overline{\cal B}} - T_{4{\cal B} \overline{\cal
B}}$, respectively. We should have
\begin{equation}
T^{(\prime)}_{1{\cal B}\overline{\cal D}}
 =\frac{1}{2}T^{(\prime)}_{2{\cal B}\overline{\cal D}},
\label{eq:BDLambda}
\end{equation}
and
 \begin{equation}
 T^{(\prime)}_{1{\cal B} \overline{\cal B}}
 =\frac{1}{2} T^{(\prime)}_{3{\cal B} \overline{\cal B}},
 \quad
 T^{(\prime)}_{2{\cal B} \overline{\cal B}}
 =\frac{1}{2} T^{(\prime)}_{4{\cal B} \overline{\cal B}}.
 \label{eq:BBLambda}
\end{equation}
Although above relations are obtained in the $\Delta S=0$ case, as
shown in the appendix, they also hold in the $|\Delta S|=1$ case.

Furthermore, we note that there should be no penguin contribution
to the $\overline B {}^0\to \Lambda \overline {\Lambda}$ decay
amplitude.
In the penguin type $b\to d_L$ decay, the quark content of the
produced $\Lambda$ should be $d_L u_R s_L$, since the $u$ $d$ pair
of the $\Lambda$ wave function is in a spin zero configuration.
From the helicity (or chirality) conservation, the correlated
anti-baryon is in a left-handed helicity state with quark content
$\bar d_L \bar u_L \bar s_R$, which cannot match to the wave
function of $\overline \Lambda$ and resulting a vanishing penguin
contribution on the $\overline B {}^0\to
\Lambda\overline{\Lambda}$ mode.  As shown in Table~\ref{tab:BB0},
its penguin amplitude is proportional to $P_{1{\cal B}
\overline{\cal B}}-5 P_{2{\cal B} \overline{\cal B}}$ and we
should have
\begin{equation}
 P^{(\prime)}_{1{\cal B} \overline{\cal B}}
 =5 P^{(\prime)}_{2{\cal B} \overline{\cal B}}.
 \label{eq:BBLL}
\end{equation}
According to the decomposition shown in Eq.~(\ref{eq:DFS}), this
relation can be expressed by $(P_D/P_F)\to (3/2)$, which is
consistent with the asymptotic relation on the baryon scalar and
pseudoscalar form factors obtained in Ref.~\cite{Chua:2002yd}.

There is no constraint from processes involving $\overline
\Lambda$ final states in the $\overline B\to {\cal
D}\overline{\cal B}$ decays and we still have two independent tree
amplitudes in $\overline B\to{\cal B}\overline {\cal B}$ decays.
Asymptotic relations lead to further reduction,
\begin{equation}
T^{(\prime}_{1{\cal D}\overline {\cal B}}=-T^{(\prime)}_{2{\cal D}
\overline {\cal B}}. \label{eq:DBasym}
\end{equation}
and
\begin{equation} T^{(\prime)}_{1{\cal B} \overline{\cal
B}}=-T^{(\prime)}_{2{\cal B} \overline{\cal B}},
 \label{eq:BBasym}
\end{equation}
in the large $m_B$ limit.

It is interesting to note that in Ref.~\cite{Brodsky:1980sx} the
octet-octet and octet-decuplet systems are related asymptotically.
Similarly, we have
 \begin{eqnarray}
 T^{(\prime)}
 =T^{(\prime)}_{{\cal D}\overline{\cal D}}
 =T^{(\prime)}_{1{\cal B}\overline{\cal D}}
  =T^{(\prime)}_{1{\cal D} \overline {\cal B}}
  =T^{(\prime)}_{1{\cal B} \overline{\cal B}},
\nonumber\\
 P^{(\prime)}
 =P^{(\prime)}_{{\cal D} \overline{\cal D}}
 =P^{(\prime)}_{{\cal B} \overline{\cal D}}
 = P^{(\prime)}_{{\cal D} \overline{\cal B}}
 =P^{(\prime)}_{1{\cal B} \overline{\cal B}},
\label{eq:asymptoticrelations}
\end{eqnarray}
in the large $m_B$ limit.
In that limit, we need only one tree and one penguin amplitudes
under the quark diagram approach for all four classes of charmless
two-body baryonic modes. In the next section, we will use the
above equation as an approximation to study the decay rates. It
should be verified by data that whether this is a good
approximation or not.

\section{\label{sec:ph}Phenomenological Discussion}

In this section, we discuss the phenomenology of charmless
two-body baryonic modes by using the formalism developed in
previous sections. Since none of the charmless two-body baryonic
decay is observed so far, we suggest some prominent modes for
experimental searches. This section is divided into two parts,
discussing tree-dominated and penguin-dominated modes,
respectively. We summarize our results at the end of this section.

\subsection{Tree-dominated modes}

\begin{table*}[t!]
\caption{\label{tab:tree} Decay rates for $\Delta S=0$
tree-dominated modes. We consider only tree amplitude
contribution. Rates are normalized to ${\mathcal B}(\overline B
{}^0\to p\overline p)$. For comparison, we show results of pole
model~\cite{Cheng:2001tr}, diquark mode~\cite{Chang:2001jt} and
sum rule~\cite{Chernyak:ag} calculations in the subsiding order in
parentheses.}
\begin{ruledtabular}
\begin{tabular}{rrlrrlrc}
Mode
          & ${\mathcal B}(10^{-7})\hspace{-1.5cm}$
          &
          & Mode
          & ${\mathcal B}(10^{-7})\hspace{-1.5cm}$
          &
          & Mode
          & ${\mathcal B}(10^{-7})$
          \\
\hline $\overline B {}^0\to p \overline{p}$
          & $1$~\footnotemark[1]
          & (1.1, 0.84, \,\, 1)
          & $B^-\to n \overline{p}$
          & $1.09$ &(5.0,\,\,\,\,\, 0, \,\,\,0.6)
          & $\overline B {}^0_s\to p \overline{\Sigma^{+}}$
          & $0.96$
          \\
$n \overline{n}$
          & $4.00$ &(1.2, 0.84, 0.3)
          & $\Sigma^0 \overline{\Sigma^+}$
          & $1.99$ &
          & $n \overline{\Lambda}$
          & $1.46$
          \\
$\Lambda\overline{\Lambda}$
          & $0$ &(0, \,\,\,\,0.38,\,\, ---)
          & $p \overline{\Delta^{++}}$
          & $6.19$ &(14, 0.63, 0.25)
          & $n \overline{\Sigma^{0}}$
          & $0.48$
          \\
$\Sigma^{0} \overline{\Lambda}$
          & $2.79$ &
          & $n \overline{\Delta^+}$
          & $2.06$ &(4.6, 0.70,\,\,\, ---)
          & $\Sigma^0 \overline{\Xi^0}$
          & $7.17$
          \\
$\Sigma^{0} \overline{\Sigma^0}$
          & $0.91$ &
          & $\Sigma^0 \overline{\Sigma^{*+}}$
          & $3.81$ &
          & $p \overline{\Sigma^{*+}}$
          & $1.85$
          \\
$p \overline{\Delta^+}$
          & $1.90$ &(4.3, 0.28, 0.1)
          & $\Delta^0 \overline{p}$
          & $2.06$ &
          & $n \overline{\Sigma^{*0}}$
          & $0.92$
          \\
$n \overline{\Delta^0}$
          & $1.90$ &(4.3, 0.28,\,\, ---)
          & $\Sigma^{*0} \overline{\Sigma^{+}}$
          & $0.95$ &
          & $\Sigma^{0} \overline{\Xi^{*0}}$
          & $3.40$
          \\
$\Sigma^{0} \overline{\Sigma^{*0}}$
          & $1.75$ &
          & $\Delta^{+} \overline{\Delta^{++}}$
          & $11.72$ &
          & $\Delta^+ \overline{\Sigma^{+}}$
          & $1.83$
          \\
$\Delta^+ \overline{p}$
          & $1.90$ &
          & $\Delta^0 \overline{\Delta^+}$
          & $3.91$ &
          & $\Delta^0\overline{\Lambda}$
          & $2.78$
          \\
$\Delta^0 \overline{n}$
          & $7.60$ &
          & $\Sigma^{*0} \overline{\Sigma^{*+}}$
          & $1.82$ &
          & $\Delta^0 \overline{\Sigma^{0}}$
          & $0.91$
          \\
$\Sigma^{*0} \overline{\Lambda}$
          & $1.34$ &
          &
          &        &
          & $\Sigma^{*0} \overline{\Xi^{0}}$
          & $3.44$
          \\
$\Sigma^{*0} \overline{\Sigma^{0}}$
          & $0.44$ &
          &
          &        &
          & $\Delta^{+} \overline{\Sigma^{*+}}$
          & $3.50$
          \\
$\Delta^+ \overline{\Delta^{+}}$
          & $3.60$ &
          &
          &        &
          & $\Delta^0 \overline{\Sigma^{*0}}$
          & $1.75$
          \\
$\Delta^0 \overline{\Delta^0}$
          & $3.60$ &
          &
          &        &
          & $\Sigma^{*0} \overline{\Xi^{*0}}$
          & $1.63$
          \\
$\Sigma^{*0} \overline{\Sigma^{*0}}$
          & $0.84$ &
          &
          &        &
          &
          \\
\end{tabular}
\end{ruledtabular}
\footnotetext[1] {We take ${\mathcal B}(\overline B {}^0\to
p\overline p)=1\times 10^{-7}$ as a reference rate.}

\end{table*}

For $\Delta S=0$ modes, we expect tree amplitudes to dominate. We
can estimate their relative rates by neglecting penguin
contribution. Rates are normalized to the $\overline B {}^0\to
p\overline p$ rate.
As noted before, a simple scaling of $|V_{ub}/V_{cb}|^2$ on the
$\overline B {}^0\to\Lambda_c^+\overline p$ decay rate hints at a
$\sim 10^{-7}$ rate for the charmless case~\cite{Gabyshev:2002dt}.
A pole model calculation also gives~${\mathcal B}(\overline B
{}^0\to p\overline p)=1.1\times 10^{-7}$~\cite{Cheng:2001tr}. For
illustration, we take ${\mathcal B}(\overline B {}^0\to p\overline
p)=1\times 10^{-7}$ as the reference rate for these tree-dominated
decay rates.

As noted in the end of the previous section, we use the asymptotic
relations stated in
Eqs.~(\ref{eq:BDLambda}--\ref{eq:asymptoticrelations})
for tree and penguin amplitudes shown in
Tables~\ref{tab:DD}--\ref{tab:BB0}.
Hadron masses and $B^{-,0}$, $B_s$ lifetimes taken from
Ref.~\cite{PDG} are used 
to indicate some SU(3) breaking effects. Results obtained are
given in Table~\ref{tab:tree}. For comparison, we quote results of
pole model~\cite{Cheng:2001tr}, diquark mode~\cite{Chang:2001jt}
and sum rule~\cite{Chernyak:ag} calculations in the subsiding
order in parentheses.
We normalize the tree contributed ${\mathcal B}_{\rm T}(\overline
B {}^0 \to p\overline p)$ of Ref.~\cite{Cheng:2001tr} and
Ref.~\cite{Chernyak:ag} to $1\times 10^{-7}$. Note that the
$\overline B {}^0\to p\overline p$ rate only decreased by 16\% by
including penguin contributions in Ref.~\cite{Chang:2001jt}. The
tree-penguin interference effect is not prominent. We will wait
for data
before carry out a detail study of the tree-penguin interference
effect.

Rate relations shown in Table~\ref{tab:tree} are consistent with
Eqs.~(\ref{eq:relDD0}, \ref{eq:relBD0}, \ref{eq:relDB0},
\ref{eq:relDB1}, \ref{eq:relBB0}).
We find that the tree-dominated $\overline B\to
N\overline{\Delta}$ modes have the structure
\begin{eqnarray}
 {\mathcal B}(B^-\to p\overline{\Delta^{++}})
 &\simeq&3\,{\mathcal B}(\overline B {}^0\to p\overline{\Delta^+})
 \nonumber\\
 &\simeq&3\,{\mathcal B}(B^-\to n\overline{\Delta^+})
 \simeq3\,{\mathcal B}(\overline B {}^0\to n\overline{\Delta^0}),
 \nonumber\\
\end{eqnarray}
which is consistent with Refs.~\cite{Jarfi:1990ej,Cheng:2001tr},
while the ${\mathcal B}(\overline B\to
N\overline{\Delta})/{\mathcal B}(\overline B {}^0\to
p\overline{p})$ ratios are smaller than those obtained in
Ref.~\cite{Cheng:2001tr} by a factor of two. We have ${\mathcal
B}(\overline B {}^0\to p\overline p)\simeq {\mathcal B}(B^-\to
p\overline{n})$, which is close to the sum-rule
result~\cite{Chernyak:ag}. On the other hand, our prediction on
${\mathcal B}(\overline B {}^0\to n\overline n):{\mathcal
B}(\overline B {}^0\to p\overline p)\simeq 4:1$ is different from
all quoted earlier results.
This different may be related to the following fact. It is easy to
see from Table~\ref{tab:BB0} and Eqs.~(\ref{eq:BBLambda},
\ref{eq:BBasym}) that by neglecting the sub-leading penguin
contribution we have
\begin{equation}
A(B^-\to n\overline p)-A(\overline B {}^0\to p\overline p)=
A(\overline B {}^0\to n\overline n). \label{eq:NNbar}
\end{equation}
The origin of the above equation is the corresponding $e_T$
coefficients shown in Table~\ref{tab:eTeP}. Note that the above
equation is different from the $\Delta I=\frac{1}{2}$ rule
relation~\cite{Jarfi:1990ej,Cheng:2001tr}, which implies
$|A(\overline B {}^0\to p\overline p)-A(\overline B {}^0\to
n\overline n)|= |A(B^-\to n\overline p)|$. In fact, a highly
suppressed ${\mathcal B}(B^-\to \pi^-\pi^0)/{\mathcal B}(\overline
B {}^0\to \pi^+\pi^-)$ ratio is followed from the $\Delta
I=\frac{1}{2}$ rule in the charmless mesonic sector. Although the
present data is still fluctuating, it suggests ${\mathcal
B}(\pi^-\pi^0)/{\mathcal B}(\pi^+\pi^-)\sim 1$~\cite{pipi0}. We do
not worry about the deviation of the $\Delta I=\frac{1}{2}$ rule
in Eq.~(\ref{eq:NNbar}) at this stage. Note that we have
${\mathcal B}(\overline B {}^0\to\Lambda\overline \Lambda)=0$. As
shown in the previous section, the vanishing of the $\overline B
{}^0\to \Lambda\overline \Lambda$ rate is due to the
anti-symmetric property of the $\Lambda$ wave function and the
chirality argument in the large $m_B$ limit. This mode provides a
useful check of the approximation made.

Upper limits on $\overline B {}^0\to p\overline p$,
$\Lambda\overline\Lambda$ and $B^-\to\Lambda\overline p$ are
obtained~\cite{Abe:2002er} (see, Eq.~(\ref{eq:UL})). Some observed
three-body modes~\cite{Abe:2002ds,Wang:2003zy,Wang:2003yi} also
provide constraints. To discuss the feasibility of experimental
observation, it is useful to recall that
 ${\mathcal B}(\Sigma^0\to\Lambda\gamma)\simeq 1$,
 ${\mathcal B}(\Xi\to\Lambda\pi)\sim 1$,
 ${\mathcal B}(\Delta^+\to p\pi^0)\simeq\frac{2}{3}$,
 ${\mathcal B}(\Delta^0\to p\pi^-)\simeq\frac{1}{3}$ and
 ${\mathcal B}(\Sigma^{*}\to \Lambda\pi)=88\%$~\cite{PDG}.

As shown in the first column of Table~\ref{tab:tree}, there are
ten $\overline B {}^0$ decay modes having rates larger than the
reference $\overline B {}^0\to p\overline p$ rate. We can search
for the $\Sigma^0\overline{\Lambda}$ mode, which has a rate about
three times of the $p\overline{p}$ rate and clear signature for
reconstruction. Up to now, most experimental searches in baryonic
modes do not involve $\pi^0$ in reconstruction. With one $\pi^0$
in the final states, we can search for $p\overline{\Delta^+}$,
$\Delta^+ \overline p$, $\Sigma^0\overline{\Sigma^{*0}}$ and
$\Sigma^{*0}\overline \Lambda$ modes, which have rates about twice
of the $\overline B {}^0\to p\overline p$ rate.
Note that $\overline B {}^0\to n\overline n$, $\Delta^0\overline
n$ decays have rates larger than the $p\overline p$ rate by
factors of four and eight, respectively, but require $\overline n$
or even $n$ in detection.

We now turn to $B^-$ decay modes as shown in the second column of
Table~\ref{tab:tree}.
One should search for the $\Delta^+\overline{\Delta^{++}}$ mode,
which have the largest rate, ${\mathcal B}(B^-\to
\Delta^+\overline{\Delta^{++}})=11.7\,{\mathcal B}(\overline B
{}^0\to p\bar p)$, and only reduces $\sim30\%$ in producing
$p\,\pi^0\,\overline p\,\pi^-$ final state.
Two modes, $p\overline{\Delta^{++}}$ and $\Delta^0\overline p$,
can decay to the all charged $p\,\overline p\,\pi^-$ final state.
They have rates six and two times of ${\mathcal B}(\overline B
{}^0\to p\bar p)$ and are searchable. Their rates are within the
tree-body (total) rate constraint, ${\mathcal B}(p\overline
p\pi^-)=(3.06^{+0.73}_{-0.62}\pm0.40)\times10^{-6}$~\cite{Wang:2003zy},
and should be accessible.

For the $B_s$ case, one needs all charge track for detection. As
shown in the third column of Table~\ref{tab:tree} the
$p\overline{\Sigma^{(*)+}}$ and $\Delta^0\overline\Lambda$ modes
have rates of one to three times of ${\mathcal B}(p\bar p)$ and
should be searchable.


There are some pure-tree modes in $|\Delta S|=1$ processes. We use
$T^\prime=T\, V^*_{us}/V^*_{ud}$ to estimate their rates and
obtain ${\mathcal B}( B^-\to
\Lambda\overline{\Delta^+})=0.15\times10^{-7}$ and ${\mathcal
B}(\overline  B {}^0_s\to
\Lambda\overline{\Sigma^{*0}},\,\Sigma^{*0}\overline{\Lambda})=0.07\times10^{-7}$,
which are quite small as expected.
There are some pure-penguin modes in the $\Delta S=0$ process as
shown in Tables~\ref{tab:DD}--\ref{tab:BB0}. Their rates are
suppressed and will be briefly discussed in the next sub-section.

\subsection{Penguin-dominated modes}

\begin{table*}[t!]
\caption{\label{tab:penguin} Decay rates for $|\Delta S|=1$
penguin-dominated modes. We consider only penguin amplitude
contribution. Rates are normalize to ${\mathcal B}(B^-\to \Lambda
\overline p)$. For comparison, we show results of pole
model~\cite{Cheng:2001tr}, diquark mode~\cite{Chang:2001jt} and
sum rule~\cite{Chernyak:ag} calculations in the subsiding order in
parentheses.}
\begin{ruledtabular}
\begin{tabular}{rrlrrlrc}
Mode
          & ${\mathcal B}(10^{-7})\hspace{-1.5cm}$
          &
          & Mode
          & ${\mathcal B}(10^{-7})\hspace{-1.5cm}$
          &
          & Mode
          & ${\mathcal B}(10^{-7})$
          \\
\hline $B^-\to \Lambda \overline{p}$
          & $1$~\footnotemark[2] &(2.2, 0.18, $<3.8$)
          &$\overline B {}^0\to \Sigma^{+}\overline{p}$
          & $0.07$               &(0.2, 0.88, 7.5)
          &$\overline B {}^0_s\to \Sigma^+ \overline{\Sigma^+}$
          & $0.06$
          \\
$\Sigma^0\overline{p}$
          & $0.04$                  &(0.6, 0.18,\,\,\,\,\, 3.8)
          & $\Lambda \overline{n}$
          & $0.92$~\footnotemark[3] &(2.1, 0.21,\, ---)
          & $\Lambda \overline{\Lambda}$
          & $0.60$
          \\
$\Xi^0 \overline{\Sigma^+}$
          & $1.70$                  &
          & $\Sigma^{0} \overline{n}$
          & $0.03$~\footnotemark[3] &(\,---\,, 0.21,\, ---)
          & $\Sigma^{0} \overline{\Sigma^0}$
          & $0.06$
          \\
$\Sigma^{0} \overline{\Delta^+}$
          & $0.28$                  &(\,---\,, 0.08,\,\,\,\,\,\, ---)
          & $\Xi^0\overline{\Sigma^0}$
          & $0.78$                  &
          & $\Xi^{0} \overline{\Xi^0}$
          & $0.98$
          \\
$\Xi^{0} \overline{\Sigma^{*+}}$
          & $0.13$                  &
          & $\Sigma^{0} \overline{\Delta^0}$
          & $0.26$                  &(\,---\,, 0.17,\, ---)
          & $\Sigma^{+} \overline{\Sigma^{*+}}$
          & $0.12$
          \\
$\Sigma^{+} \overline{\Delta^{++}}$
          & $0.42$                  &(2.0, 1.10,\,\,\,\,\, 7.5)
          & $\Xi^{0} \overline{\Sigma^{*0}}$
          & $0.06$                  &
          & $\Sigma^{0} \overline{\Sigma^{*0}}$
          & $0.12$
          \\
$\Sigma^{*0} \overline{p}$
          & $0.07$                  &
          & $\Sigma^{+} \overline{\Delta^{+}}$
          & $0.13$                  &(0.6, 0.57, 7.5)
          & $\Xi^0 \overline{\Xi^{*0}}$
          & $0.12$
          \\
$\Xi^{*0} \overline{\Sigma^{+}}$
          & $0.13$                  &
          & $\Sigma^{*0} \overline{n}$
          & $0.06$                  &
          & $\Sigma^{*0} \overline{\Sigma^{0}}$
          & $0.12$
          \\
$\Sigma^{*0} \overline{\Delta^+}$
          & $0.53$                  &
          & $\Sigma^{*+} \overline{p}$
          & $0.13$                  &
          & $\Xi^{*0} \overline{\Xi^{0}}$
          & $0.12$
          \\
$\Sigma^{*+} \overline{\Delta^{++}}$
          & $0.80$                  &
          & $\Xi^{*0} \overline{\Lambda}$
          & $0.20$                  &
          & $\Sigma^{*+} \overline{\Sigma^{+}}$
          & $0.12$
          \\
$\Xi^{*0} \overline{\Sigma^{*+}}$
          & $0.98$                  &
          & $\Xi^{*0} \overline{\Sigma^{0}}$
          & $0.06$                  &
          & $\Xi^{*0} \overline{\Xi^{*0}}$
          & $0.88$
          \\
          &                         &
          & $\Sigma^{*0} \overline{\Delta^0}$
          & $0.49$                  &
          & $\Sigma^{*0} \overline{\Sigma^{*0}}$
          & $0.24$
          \\
          &                         &
          & $\Sigma^{*+} \overline{\Delta^{+}}$
          & $0.24$                  &
          & $\Sigma^{*+} \overline{\Sigma^{*+}}$
          & $0.24$
          \\
          &                         &
          & $\Xi^{*0} \overline{\Sigma^{*0}}$
          & $0.45$                  &
          \\
\end{tabular}
\end{ruledtabular}
\footnotetext[2]{We take ${\mathcal B}(B^-\to \Lambda \overline
p)=1\times 10^{-7}$ as our reference rate.}
\footnotetext[3]{These modes may have large tree contributions.
See text for detail discussion.}
\end{table*}

\begin{table*}[t!]
\caption{\label{tab:purepenguin} Rates (normalized to ${\mathcal
B}(B^-\to \Lambda \overline p)=1\times 10^{-7}$) for pure penguin
modes in the $|\Delta S|=1$ process. For comparison we show
results of Ref.~\cite{Cheng:2001tr} in the parenthesis.}
\begin{ruledtabular}
\begin{tabular}{rlrcrc}
Mode
          & ${\mathcal B}(10^{-7})$
          & Mode
          & ${\mathcal B}(10^{-7})$
          & Mode
          & ${\mathcal B}(10^{-7})$
          \\
\hline $B^-\to \Xi^{-} \overline{\Sigma^0}$
          & $0.85$
          & $\overline B {}^0\to \Xi^- \overline{\Sigma^-}$
          & $1.56$
          & $\overline B {}^0_s\to \Xi^{-} \overline{\Xi^{-}}$
          & $0.98$
          \\
$\Xi^{-} \overline{\Lambda}$
          & $0.10$
          & $\Xi^{0} \overline{\Lambda}$
          & $0.10$
          & $\Sigma^{-} \overline{\Sigma^{-}}$
          & $0.06$
          \\
$\Sigma^{-} \overline{n}$
          & $0.07$
          & $\Xi^{-} \overline{\Sigma^{*-}}$
          & $0.12$
          & $\Sigma^{-} \overline{\Sigma^{*-}}$
          & $0.12$
          \\
$\Xi^-\overline{\Sigma^{*0}}$
          & $0.06$
          & $\Sigma^{-} \overline{\Delta^{-}}$
          & $0.38$
          & $\Xi^{-} \overline{\Xi^{*-}}$
          & $0.12$
          \\
$\Sigma^{-} \overline{\Delta^{0}}$
          & $0.14$\,\,\,\,\,  (0.7)
          & $\Xi^{*-} \overline{\Sigma^{-}}$
          & $0.12$
          & $\Sigma^{*-} \overline{\Sigma^{-}}$
          & $0.12$
          \\
$\Sigma^{*-} \overline{n}$
          & $0.14$
          & $\Omega^{-} \overline{\Xi^{-}}$
          & $0.33$
          & $\Xi^{*-} \overline{\Xi^{-}}$
          & $0.12$
          \\
$\Xi^{*-} \overline{\Lambda}$
          & $0.20$
          & $\Sigma^{*-} \overline{\Delta^-}$
          & $0.73$
          & $\Sigma^{*-} \overline{\Sigma^{*-}}$
          & $0.24$
          \\
$\Xi^{*-} \overline{\Sigma^{0}}$
          & $0.06$
          & $\Xi^{*-} \overline{\Sigma^{*-}}$
          & $0.90$
          & $\Xi^{*-} \overline{\Xi^{*-}}$
          & $0.88$
          \\
$\Omega^{-} \overline{\Xi^{0}}$
          & $0.36$
          & $\Omega^{-} \overline{\Xi^{*-}}$
          & $0.62$
          & $\Omega^{-} \overline{\Omega^{-}}$
          & $1.82$
          \\
$\Sigma^{*-} \overline{\Delta^0}$
          & $0.27$
          &
          &
          &
          \\
$\Xi^{*-} \overline{\Sigma^{*0}}$
          & $0.49$
          &
          &
          &
          &
          \\
$\Omega^{-} \overline{\Xi^{*0}}$
          & $0.68$
          \\
\end{tabular}
\end{ruledtabular}
\end{table*}

The relative rates for penguin-dominated modes can be obtained
similarly as in the previous subsection, but instead of neglecting
penguin amplitudes we neglect tree amplitudes. We use
$B^-\to\Lambda\overline p$ as a reference mode. The present
experimental limit is ${\mathcal B}(B^-\to\Lambda\overline
p)<2.2\times 10^{-6}$~\cite{Abe:2002er}. On the theoretical side,
a recent pole model calculation gives ${\mathcal
B}(B^-\to\Lambda\overline p)=2\,{\mathcal B}(\overline B {}^0\to
p\overline p)=2.2\times 10^{-7}$~\cite{Cheng:2001tr}, while a
diquark model calculation gives ${\mathcal
B}(B^-\to\Lambda\overline p)=0.18\,{\mathcal B}(\overline B
{}^0\to p\overline p)$~\cite{Chang:2001jt}. Due to the
controversial situation of these predictions, we use ${\mathcal
B}(B^-\to\Lambda\overline p)=1\times 10^{-7}$ as a convenient
reference rate for illustration. In Table~\ref{tab:penguin} we
show rates for penguin-dominated modes, while rates for
pure-penguin modes are given in Table~\ref{tab:purepenguin}. Rate
ratios are consistent with Eqs.~(\ref{eq:relDD1}, \ref{eq:relBD1},
\ref{eq:relDB1}, \ref{eq:relBB1}, \ref{eq:relBB2}).
For comparison, we show results of pole model~\cite{Cheng:2001tr},
diquark mode~\cite{Chang:2001jt} and sum rule~\cite{Chernyak:ag}
calculations, in the subsiding order in parentheses. The diquark
and sum rule results are normalized as before.

The ${\mathcal B}(\overline B {}^0\to \Lambda\overline
n)/{\mathcal B}(B^-\to \Lambda\overline p)$ ratio is similar to
those obtained in pole model and diquark model
calculations~\cite{Cheng:2001tr,Chang:2001jt}, while the
${\mathcal B}(\overline B {}^0\to \Sigma^+\overline p)/{\mathcal
B}(B^-\to \Lambda\overline p)$ ratio is consistent with the pole
model result~\cite{Cheng:2001tr}. For other cases, we have quite
different predictions.

As shown in Table~\ref{tab:penguin}, the $B^-\to\Lambda\overline
p$ decay is still the best mode to search for. Other modes may
have much smaller rates or much lower detection efficiencies. For
example, although the $B^-\to \Sigma^0\overline p$ decay signature
is easy to identify, its rate is too small to search for. This
smallness as compared to the $\Lambda\overline p$ rate can be
traced to the $e_P$ ratio ($-1/3\sqrt2:\sqrt{3/2}$) as shown in
Table~\ref{tab:eTeP}. Note that the same ratio is obtained in the
$\Lambda \overline p$ vs. $\Sigma^0\overline p$ (pseudo)scalar
form factors in the asymptotic limit~\cite{Chua:2002yd}. The
resulting prediction , ${\mathcal B}(\Lambda\overline
p\pi^-)>{\mathcal B}(\Sigma^0\overline p\pi^-)$, is supported by
data~\cite{Wang:2003yi}.
%
The $B^-\to\Xi^0\overline{\Sigma^+}$ rate is about twice the
$\Lambda\overline p$ rate, but we need two $\pi^0$ in
reconstruction. Note that an all charged final state can be found
in the
$B^-\to\Sigma^{*+}\overline{\Delta^{++}}\to\Lambda\pi^+\overline
p\,\pi^-$ decay with only 10\% reduction in rate.

The penguin-dominated $\overline B {}^0$ decay rates are shown in
the second column of Table~\ref{tab:penguin}. In general their
rate are within ${\mathcal B}(B^-\to\Lambda\overline p)$. We note
that the $\Sigma^0\overline{\Delta^0}$ final state has one third
of chance to decay to $\Sigma^0\overline{p}\,\pi^-$ and the rate
is within the three-body total rate constraint (${\mathcal
B}(\Sigma^0\overline p\pi^-)<3.8\times 10^{-6}$ at 90\% confidence
level~\cite{Wang:2003yi}). For the $B_s$ case, the
$\Lambda\overline{\Lambda}$ mode is the most searchable one.

We note that some modes may have large tree contributions. For
example, as one can see from Table~\ref{tab:BB1} and
Eqs.~(\ref{eq:BBLambda}, \ref{eq:BBasym}) that $\overline B
{}^0\to \Lambda\overline{n} ,\,\Sigma^0\overline{n}$ amplitudes
have relatively large $T^\prime$ coefficients. By using
$T^\prime=T\, V^*_{us}/V^*_{ud}$, we have ${\mathcal B}_{\rm
T}(\overline B {}^0\to \Lambda\overline{n}
,\,\Sigma^0\overline{n}) =(0.12,\,0.07) \,{\mathcal B}(\overline B
{}^0\to p\overline p)$ from tree contributions. For these modes,
tree-penguin interference may affects their rates dramatically.
For example, by including tree contribution in $\overline B
{}^0\to\Sigma^0\overline{n}$ decay, Ref.~\cite{Chang:2001jt}
obtains a factor of five enhancement of an earlier
calculation~\cite{Ball:1990fw}.
%
As one can see from Tables~\ref{tab:DD}--\ref{tab:BB1} and
asymptotic relations that other $|\Delta S|=1$ modes
do not have such a large relative tree
contribution. The tree-penguin interference for them are expected
to be much milder.

Rates for pure-penguin modes are shown in
Table~\ref{tab:purepenguin}. They are in general not favorable
than the $B^-\to\Lambda\overline p$ mode. The largest mode of
pure-penguin $B^-$ decays is $\Xi^-\overline{\Sigma^0}$ and one
may search for it after the observation of the $\Lambda\overline
p$ mode. Although the $\overline B
{}^0\to\Xi^-\overline{\Sigma^-}$ rate is 1.6 times of
$\Lambda\overline p$ rate, we need $\Lambda\pi^- \overline
n\,\pi^+$ for detection. One may have chance to search for
$\overline B {}^0_s\to\Xi^-\overline{\Xi^-}$,
$\Omega^-\overline{\Omega^-}$ decays through
$\Lambda\pi^-\overline\Lambda\pi^+$, $\Lambda K^-\overline\Lambda
K^+$ final states, respectively.

As noted in the end of the previous subsection, there are some
pure penguin decays in $\Delta S=0$ processes. We can estimate
their rate by using $P=P^\prime\,V_{td}^*/V_{ts}^*$. Their rates
are typically of $10^{-9}$ and even the largest rate, ${\mathcal
B}(\overline B {}^0\to \Delta^-\overline{\Delta^-})\simeq 9\times
10^{-9}$, is still undetectable.

We give a summary of our suggestions before ending this section.
We find that in $\Delta S=0$ processes, in addition of the
$\overline B {}^0\to p\overline p$ search, it is useful to search
for the $\overline B {}^0\to\Sigma^0\overline \Lambda$ decay and
$B^-\to\Delta^+\overline{\Delta^{++}}$, $p\overline{\Delta^{++}}$,
$\Delta^0\overline p$ decays. In particular, the
$\Delta^+\overline{\Delta^{++}}$ rate is larger than the $p\bar p$
rate by one order of magnitude. It could be the first charmless
two-body baryonic mode to appear.
For $|\Delta S|=1$ processes, the $B^-\to\Lambda\overline p$ decay
is still the best mode to search for. After the observation of any
of the above mention modes, one should also search for other
sub-dominated modes, such as $\overline B {}^0\to
p\overline{\Delta^+}$, $\Delta^+ \overline p$,
$\Sigma^0\overline{\Sigma^{*0}}$, $\Sigma^{*0}\overline \Lambda$,
$\Xi^-\overline{\Sigma^-}$ and $B^-\to\Xi^-\overline{\Sigma^0}$,
$\Sigma^{*+}\overline{\Delta^{++}}$ decays. For the $B_s$ case, as
all charged tracks final states are required for detection, one
can search for the $p\overline{\Sigma^{(*)+}}$,
$\Delta^0\overline\Lambda$, $\Xi^-\overline{\Xi^-}$ and
$\Omega^-\overline{\Omega^-}$ modes, which have rates compatible
to $\overline B {}^0\to p\overline p$ or $B^-\to\Lambda\overline
p$ decay rates. We note that although rates shown in
Table~\ref{tab:tree} seems more promising than those shown
Table~\ref{tab:penguin} and Table~\ref{tab:purepenguin}, the
reference $B^-\to\Lambda\overline p$ rate may be greater than the
reference $\overline B {}^0\to p\overline p$ rate and resulting a
different prospect. Nevertheless, there are many promising modes,
as good as the $p\overline p$ mode, in $\Delta S=0$ processes,
while the $\Lambda p$ mode is still be best candidate in $|\Delta
S|=1$ processes.

\section{Discussion and conclusion}

In this work, we use a quark diagram approach to classify various
contributions on charmless two-body baryonic $B$ decays. This
approach is closely related to flavor symmetry, since a quark
diagram is a representation of it. Relations on decay amplitudes
are obtained. As noted in Sec.~\ref{sec:formulation}, our results
are consistent with some generic SU(3) results with the OZI rule
imposed. Since we match baryon and $qqq$ flavors in the beginning,
the OZI rule should be satisfied. In this sense, we may view the
diagramatic approach as an flavor symmetry approach with the
OZI-rule constraint.

We concentrated on tree and strong penguin contributions in decay
rates. Other contributions are sub-leading and can be included
later. For example, the electroweak penguin contribution can be
included by a simple re-definition of tree and strong penguin
amplitudes as in the mesonic case~\cite{Gronau:1995hn}. Some
contributions, such as weak exchange and annihilation topologies,
can be included by extending Eq.~(\ref{eq:DD}) to other cases, but
they are expected to be $1/m_B^2$ suppressed.
Amplitude relations are obtained by considering tree and penguin
contributions.

There are in general more than one tree and more than one penguin
amplitudes to describe charmless two-body baryonic decays.
Reduction in the number of independent amplitudes can be achieved
by considering the chirality nature of weak interaction and the
large $m_B$ limit. Some relations can be understood by using the
anti-symmetry behavior of the $\Lambda$ wave function. An
interesting check is the experimental verification of the
vanishing or highly suppressed $\overline B
{}^0\to\Lambda\overline{\Lambda}$ rate.
As in Ref.~\cite{Brodsky:1980sx}, the octet-octet and
octet-decuplet systems are related asymptotically, we need only
one tree and one penguin amplitudes to describe charmless two-body
baryonic decays.
The followed results are very predictive and can be
checked by experiments.

We suggest some prominent modes to search for. In $\Delta S=0$
processes, there are many promising modes. In addition of the
$\overline B {}^0\to p\overline p$ search, it is useful to search
for $\overline B {}^0\to\Sigma^0\overline \Lambda$ decay and
$B^-\to\Delta^+\overline{\Delta^{++}}$, $p\overline{\Delta^{++}}$,
$\Delta^0\overline p$ decays.
%
For $|\Delta S|=1$ processes, the $B^-\to\Lambda\overline p$ decay
is still the best mode to search for.
After the observation of any of the above mention modes, one
should also search for other sub-dominated modes.
For the $B_s$ case, one can search for the
$p\overline{\Sigma^{(*)+}}$, $\Delta^0\overline\Lambda$,
$\Xi^-\overline{\Xi^-}$ and $\Omega^-\overline{\Omega^-}$ decay
modes. Although the above suggestions are obtained by considering
the dominant contributions, we do not expect large modification of
relative rate ratio in most cases. The tree-penguin interference
effects can be included later after the appearance of data.

To conclude, we use a quark diagram approach to study charmless
two-body baryonic decays. The topological amplitudes can be
extracted from data. We further apply asymptotic relations to
reduce the number of independent topological amplitudes and
obtained predictive results. We have pointed out several promising
modes to be added to the present experimental searching list. The
discovery of any one of them should be followed by a bunch of
other modes.

\begin{acknowledgments}
The author thanks H.-Y. Cheng, H.-C. Huang and H.-n. Li for
discussions. This work is supported by the National Science
Council of R.O.C. under Grant NSC-91-2811-M-002-043, the MOE CosPA
Project, and the BCP Topical Program of NCTS.
\end{acknowledgments}

\appendix
\section{Asymptotic relations in the large $m_B$ limit}

We follow Ref.~\cite{Brodsky:1980sx,Chua:2002yd} to obtain the
asymptotic relations stated in Sec.~\ref{sec:reduction}. As noted
we only need to consider helicity $\pm\frac{1}{2}$ states. The
wave function of a right-handed (helicity$=\frac{1}{2}$) baryon
can be expressed as
\begin{equation}
|{\mathbf B}\,;\uparrow\rangle\sim \frac{1}{\sqrt3}(|{\mathbf
B}\,;\uparrow\downarrow\uparrow\rangle
                +|{\mathbf B}\,;\uparrow\uparrow\downarrow\rangle
                +|{\mathbf B}\,;\downarrow\uparrow\uparrow\rangle),
\end{equation}
i.e. composed of 13-, 12- and 23-symmetric terms, respectively.
For ${\mathbf
B}=\Delta,\,\Sigma^{*+,0},\,p,\,n,\,\Sigma^0,\,\Lambda$, we have
\begin{eqnarray}
|\Delta^{++};\uparrow\downarrow\uparrow\rangle&=&u(1)u(2)u(3)|\uparrow\downarrow\uparrow\rangle,
\nonumber\\
|\Delta^{-};\uparrow\downarrow\uparrow\rangle&=&d(1)d(2)d(3)|\uparrow\downarrow\uparrow\rangle,
\nonumber\\
|\Delta^{+};\uparrow\downarrow\uparrow\rangle&=&
\frac{1}{\sqrt3}[u(1)u(2)d(3)+u(1)d(2)u(3)
\nonumber\\&&+d(1)u(2)u(3)]|\uparrow\downarrow\uparrow\rangle,
\nonumber\\
|\Delta^{0};\uparrow\downarrow\uparrow\rangle&=&(|\Delta^{+};\uparrow\downarrow\uparrow\rangle\,\,{\rm
with}\,\,u \leftrightarrow d),
 \nonumber\\
|\Sigma^{*+};\uparrow\downarrow\uparrow\rangle&=&(|\Delta^{+};\uparrow\downarrow\uparrow\rangle\,\,{\rm
with}\,\,d \leftrightarrow s),
\nonumber\\
|\Sigma^{*0};\uparrow\downarrow\uparrow\rangle&=&\frac{1}{\sqrt6}[u(1)d(2)s(3)+{\rm
permutation}]|\uparrow\downarrow\uparrow\rangle,
 \nonumber
 \end{eqnarray}
\begin{eqnarray}
 |p\,;\uparrow\downarrow\uparrow\rangle&=&
\bigg[\frac{d(1)u(3)+u(1)d(3)}{\sqrt6} u(2)
 \nonumber\\
 &&-\sqrt{\frac{2}{3}} u(1)d(2)u(3)\bigg]
|\uparrow\downarrow\uparrow\rangle,
\nonumber\\
|n\,;\uparrow\downarrow\uparrow\rangle&=&
(-|p\,;\uparrow\downarrow\uparrow\rangle \,\,{\rm with}\,\,u
\leftrightarrow d),
 \nonumber\\
|\Sigma^0\,;\uparrow\downarrow\uparrow\rangle&=&
\bigg[-\frac{u(1)d(3)+d(1)u(3)}{\sqrt3}\,s(2)
 \nonumber\\
      &&+\frac{u(2)d(3)+d(2)u(3)}{2\sqrt3}\,s(1)
\nonumber\\
      &&\,\,+\frac{u(1)d(2)+d(1)u(2)}{2\sqrt3}\,s(3)\bigg]
|\uparrow\downarrow\uparrow\rangle,
\nonumber\\
|\Lambda\,;\uparrow\downarrow\uparrow\rangle&=&
\bigg[\frac{d(2)u(3)-u(2)d(3)}{2}\,s(1)
 \nonumber\\
     && +\frac{u(1)d(2)-d(1)u(2)}{2}\,s(3)\bigg]
|\uparrow\downarrow\uparrow\rangle, \label{eq:wavefunction}
\end{eqnarray}
for the corresponding $|{\mathbf
B}\,;\uparrow\downarrow\uparrow\rangle$ parts, while the 12- and
23-symmetric parts can be obtained by permutation. To be
consistent with Eq.~(\ref{eq:octet}), our $\Lambda$ state has an
overall negative sign with respect to that of
Ref.~\cite{Brodsky:1980sx}.

Following Ref.~\cite{Brodsky:1980sx} and using the helicity
argument in Sec.~\ref{sec:ph}, asymptotically we have
\begin{eqnarray}
&&\langle {\mathbf B}(p_1)| {\cal O}|\overline B\,{\mathbf
B}^\prime(p_2)\rangle
=\bar u(p_1)\left[
                   \frac{1-\gamma_5}{2}\,F(t)\right] u(p_2),
\nonumber\\
&&F(t)=\sum_{i=T,P_L,P_R} e_{i}
               ({\mathbf B}^\prime -B-{\mathbf B})
               \,F_{i}(t),
\end{eqnarray}
where ${\cal O}$ are the operators in $H_{\rm eff}$. For
simplicity, we illustrate with the space-like case. Note that the
above equation is obtained in the large $t(=(p_1-p_2)^2)$ limit,
where we may take a large $m_B$ limit or view the $B$ meson as a
virtual particle with a large momentum. Quark mass dependent terms
behave like $m_q/\sqrt{|t|}$ and are neglected.
%
\begin{figure}[t!]
\centerline{\DESepsf(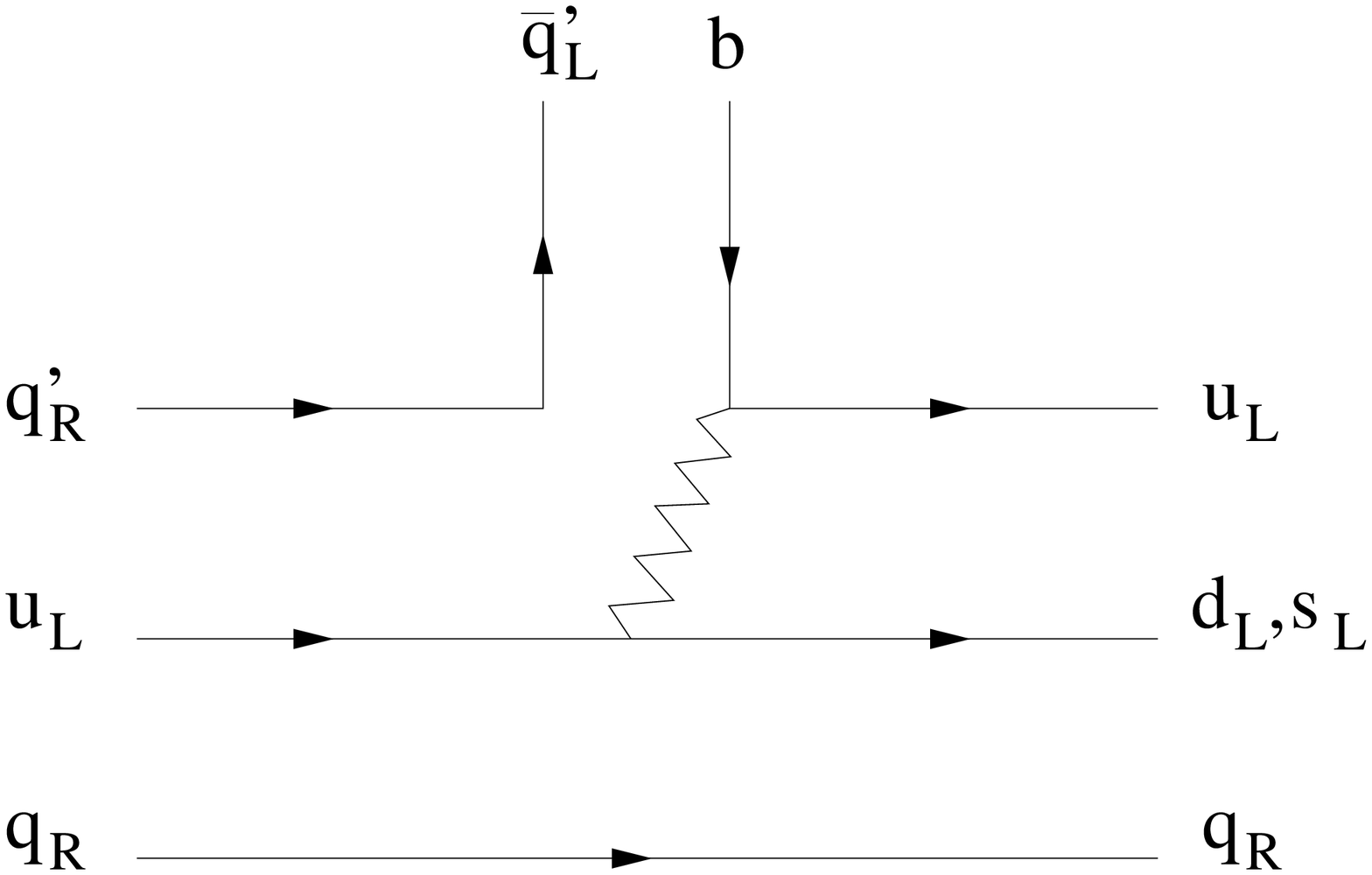 width 7cm)}
\centerline{(a)} \vskip0.2cm
\smallskip
\centerline{\DESepsf(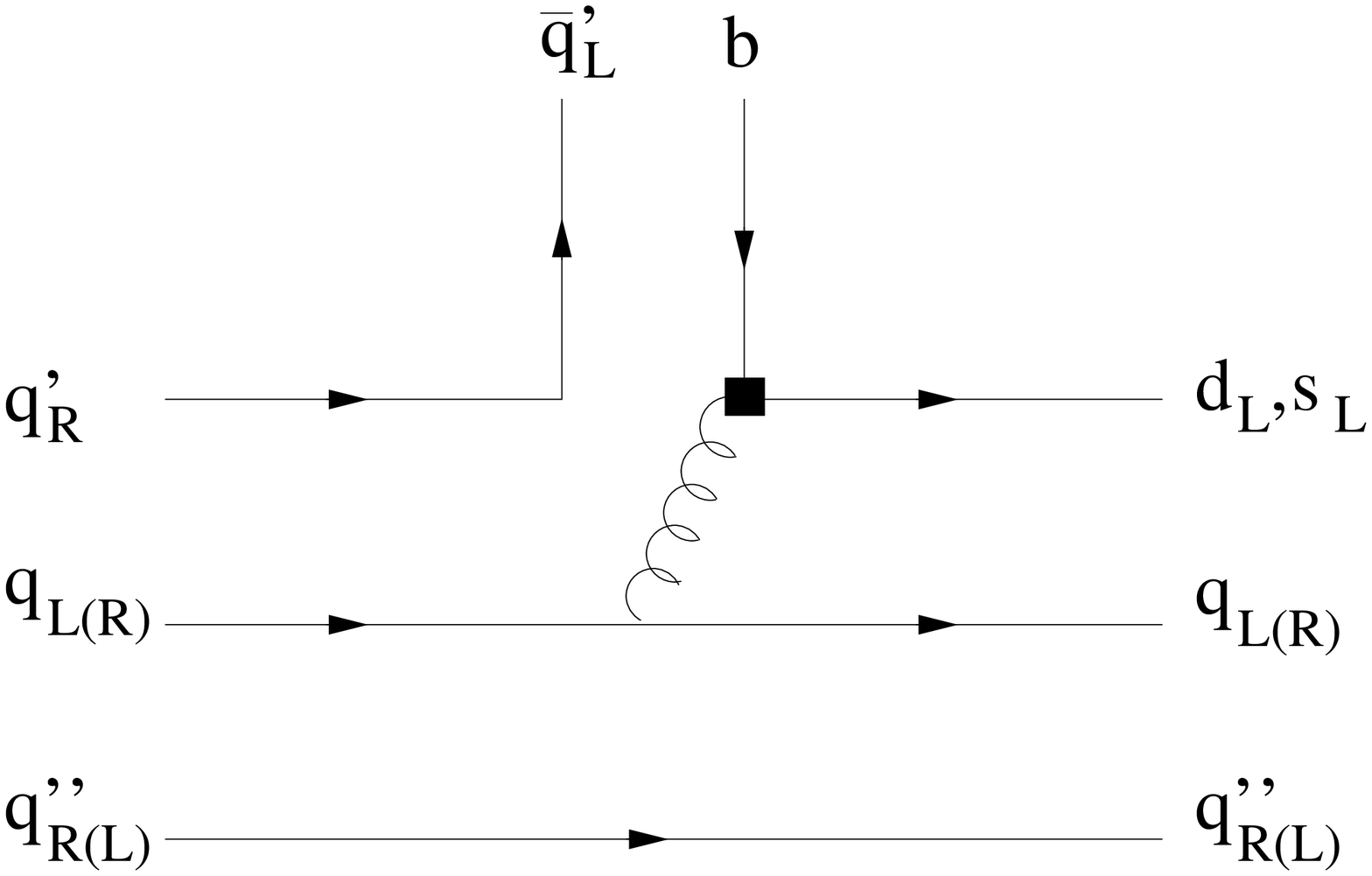 width 7cm)}
\centerline{(b)} \vskip0.2cm
\smallskip
\caption{(a) Tree and (b) penguin
  ${\mathbf B}^\prime -\overline B-{\mathbf
B}$ diagrams in the asymptotic limit. } \label{fig:eTeP}
\end{figure}
%
As shown in Fig.~\ref{fig:eTeP}(a), the tree operator $(\bar u
b)_{V-A} (\bar d u)_{V-A}$ can generate a ${\mathbf
B}^\prime(q^\prime_R\, u_L\, q_R)$--$\overline B(\overline
q^\prime_L\,b )$--${\mathbf B}(u_L \,d_L\, q_R)$ coupling. The
corresponding coefficient $e_{T}({\mathbf B}^\prime -\overline
B-{\mathbf B})$ is given by
\begin{eqnarray}
&&e_{T}({\mathbf B}^\prime -\overline B-{\mathbf B})
 \nonumber\\
&& =\langle{\mathbf B};\,\downarrow\downarrow\uparrow|
     Q[q_R^\prime(1)\to u_L(1);u_L(2)\to d_L(2)]
    |{\mathbf B}^\prime\,;\uparrow\downarrow\uparrow\rangle
\nonumber\\
&&\,\,\,+\langle {\mathbf B};\,\uparrow\downarrow\downarrow|
     Q[q_R^\prime(3)\to u_L(3);u_L(2)\to d_L(2)]
   |{\mathbf B}^\prime\,;\uparrow\downarrow\uparrow\rangle,
\nonumber\\
 \label{eq:eT}
\end{eqnarray}
where $Q[q^\prime_R(1(3))\to u_L(1,3);u_L(2)\to d_L(2)]$ changes
the $q^\prime(1(3))|\uparrow\rangle \otimes
u(2)|\downarrow\rangle$ part of $|{\mathbf
B}^\prime;\uparrow\downarrow\uparrow\rangle$ to the
$u(1(3))|\downarrow\rangle\otimes d(2)|\downarrow\rangle$ part.

\begin{table}[t!]
\caption{\label{tab:eTeP} The coefficients $e_{T,P}({\mathbf
B}^\prime -\overline B-{\mathbf B})$ for various modes obtained
from Eqs.~(\ref{eq:eT}, \ref{eq:eP}).}
\begin{ruledtabular}
\begin{tabular}{cccccc}
${\mathbf B}^\prime -\overline B-{\mathbf B}$
          & $e_T$
          & $e_P$
          & ${\mathbf B}^\prime -\overline B-{\mathbf B}$
          & $e_T$
          & $e_P$
          \\
\hline $\Sigma^{*0}-\overline B {}^0-\Sigma^{*0}$
          & $\frac{1}{3}$
          & $\frac{2}{3}$
          & $\Delta^0 -B^- -\Delta^-$
          & $0$
          & $\frac{2}{\sqrt3}$
          \\
$\Delta^+ -B^- -\Delta^0$
          & $\frac{2}{3}$
          & $\frac{4}{3}$
          & $\Delta^{+}-B^--\Sigma^{*0}$
          & $\frac{\sqrt2}{3}$
          & $\frac{2\sqrt2}{3}$
          \\
\hline $\Delta^{++}-B^- -p$
          & $\sqrt{\frac{2}{3}}$
          & $\sqrt{\frac{2}{3}}$
          & $\Sigma^{*0} -\overline B {}^0 -\Lambda$
          & $0$
          & $-\frac{1}{\sqrt6}$
          \\
$\Delta^+-\overline B {}^0-p$
          & $\frac{\sqrt2}{3}$
          & $\frac{\sqrt2}{3}$
          & $\Delta^{+}-B^- -n$
          & $-\frac{\sqrt2}{3}$
          & $\frac{\sqrt2}{3}$
          \\
$\Sigma^{*+}-\overline B {}^0_s-p$
          & $\frac{\sqrt2}{3}$
          & $\frac{\sqrt2}{3}$
          & $\Delta^{+}- B^- -\Sigma^{0}$
          & $\frac{1}{3}$
          & $\frac{2}{3}$
          \\
\hline $p-B^- -\Delta^0$
          & $\frac{\sqrt2}{3}$
          & $-\frac{\sqrt2}{3}$
          & $p-\overline B {}^0-\Delta^+$
          & $\frac{\sqrt2}{3}$
          & $\frac{\sqrt2}{3}$
          \\
$n -B^- -\Delta^-$
          & $0$
          & $-\sqrt{\frac{2}{3}}$
          & $\Lambda -\overline B {}^0 - \Sigma^{*0}$
          & $-\frac{1}{\sqrt6}$
          & $-\frac{1}{\sqrt6}$
          \\
$p -B^- -\Sigma^{*0}$
          & $\frac{1}{3}$
          & $-\frac{1}{3}$
          & $n -\overline B {}^0 -\Sigma^{*0}$
          & $\frac{2}{3}$
          & $\frac{1}{3}$
          \\
\hline $p- \overline B {}^0 - p$
          & $\frac{1}{3}$
          & $\frac{1}{3}$
          & $p- B^- - n$
          & $-\frac{1}{3}$
          & $-\frac{5}{3}$
          \\
$n- \overline B {}^0 - n$
          & $-\frac{2}{3}$
          & $-\frac{4}{3}$
          & $\Lambda- \overline B {}^0-\Sigma^0$
          & $-\frac{1}{\sqrt3}$
          & $\frac{1}{\sqrt3}$
          \\
$\Sigma^0 - \overline B {}^0 - \Sigma^0$
          & $-\frac{1}{3}$
          & $-\frac{2}{3}$
          & $\Sigma^0-\overline B {}^0-\Lambda$
          & $0$
          & $\frac{1}{\sqrt3}$
          \\
$\Lambda- \overline B {}^0 -\Lambda$
          & $0$
          & $0$
          & $n- \overline B {}^0 -\Lambda$
          & $\sqrt{\frac{2}{3}}$
          & $\sqrt{\frac{3}{2}}$
          \\
$p- B^- - \Sigma^0$
          & $\frac{1}{\sqrt{18}}$
          & $-\frac{1}{3\sqrt2}$
          & $p- B^- -\Lambda$
          & $\frac{1}{3\sqrt2}$
          & $\sqrt{\frac{3}{2}}$
          \\
\end{tabular}
\end{ruledtabular}
\end{table}

Similarly coefficients $e_{P_L,P_R}({\mathbf B}^\prime -\overline
B-{\mathbf B})$ for the ${\mathbf B}^\prime(q^\prime_R\, q_L\,
q^{\prime\prime}_R)$--$\overline B(\overline
q^\prime_L\,b)$--${\mathbf B}(d_L\, q_L\, q^{\prime\prime}_R)$ and
${\mathbf B}^\prime(q^\prime_R\, q_R
\,q^{\prime\prime}_L)$--$\overline B(\overline
q^\prime_L\,b)$--${\mathbf B}(d_L \,q_R\, q^{\prime\prime}_L)$
couplings governed respectively by the penguin operators $(\bar d
b)_{V-A} (\bar q q)_{V\mp A}$ are given by
\begin{eqnarray}
&&e_{P_L}({\mathbf B}^\prime -\overline B-{\mathbf B})
 \nonumber\\
 &&=\langle{\mathbf B};\,\downarrow\downarrow\uparrow|
     Q[q_R^\prime(1)\to d_L(1);q_L(2)\to q_L(2)]
    |{\mathbf B}^\prime\,;\uparrow\downarrow\uparrow\rangle
\nonumber\\
&&\,\,\,+\langle {\mathbf B};\,\uparrow\downarrow\downarrow|
     Q[q_R^\prime(3)\to d_L(3);q_L(2)\to q_L(2)]
   |{\mathbf B}^\prime\,;\uparrow\downarrow\uparrow\rangle,
\nonumber\\
&&e_{P}({\mathbf B}^\prime -\overline B-{\mathbf B})
 \equiv e_{P_L}({\mathbf B}^\prime -\overline B-{\mathbf B})
 \nonumber\\
 &&\qquad\qquad\qquad\quad\,\,=e_{P_R}({\mathbf B}^\prime -\overline B-{\mathbf B}).
 \label{eq:eP}
\end{eqnarray}
The corresponding diagram is shown in Fig.~\ref{fig:eTeP}(b). Note
that that $e_{P_L}$ is similar to $e_T$ with the
$q^\prime_R(1,3)\to u_L(1,3)$ and $u_L(2)\to d_L(2)$ operations
replaced by the $q^\prime_R(1,3)\to d_L(1,3)$ and $q_L(2)\to
q_L(2)$ operations, respectively. The equality of $e_{P_R}$ and
$e_{P_L}$ can be understood by interchanging $q\leftrightarrow
q^{\prime\prime}$ in ${\mathbf B}^\prime(q^\prime_R\,q_L\,
q^{\prime\prime}_R)$--$\overline B(\overline
q^\prime_L\,b)$--${\mathbf B}(d_L\, q_L\,q^{\prime\prime}_R)$ and
${\mathbf
B}^\prime(q^\prime_R\,q_R\,q^{\prime\prime}_L)$--$\overline
B(\overline q^\prime_L\,b)$--${\mathbf
B}(d_L\,q_R\,q^{\prime\prime}_L)$. The coefficients for the
$|\Delta S|=1$ case can be obtained by the suitable replacement of
$d_L\to s_L$ in the $\mathbf B$ content in Eqs.~(\ref{eq:eT},
\ref{eq:eP}).


As shown Table~\ref{tab:eTeP} the octet-octet, octet-decuplet and
decuplet-decuplet systems are related
asymptotically~\cite{Brodsky:1980sx}. With
\begin{eqnarray}
T^{(\prime)}
 \propto \frac{1}{3}\,V_{ub} V^*_{ud(s)}\,F_T\,\bar u_R v_L,
 \nonumber\\
P^{(\prime)}
 \propto \frac{1}{3}V_{tb} V^*_{td(s)}\,(F_{P_L}+\kappa^{(\prime)} F_{P_R})\,\bar u_R  v_L,
 \label{eq:asymptotic}
\end{eqnarray}
where $\kappa^{(\prime)}$ as the ratio of the corresponding Wilson
coefficients of the $(\bar d b)_{V-A} (\bar q q)_{V\pm A}$ ($(\bar
d s)_{V-A} (\bar q q)_{V\pm A}$) operators, we obtain the
asymptotic relations,
Eqs.~(\ref{eq:BDLambda}--\ref{eq:asymptoticrelations}),
by comparing Table~\ref{tab:eTeP}
with the corresponding amplitudes shown in
Tables~\ref{tab:DD}--\ref{tab:BB1}.




\end{document}